\documentclass[conference,compsoc]{IEEEtran}

\AtBeginDocument{%
  \providecommand\BibTeX{{%
    Bib\TeX}}}

\usepackage[english]{babel} % handle hyphenation
\usepackage[utf8]{inputenc}
\usepackage{blindtext}

\usepackage{algorithmic}


\usepackage{amsmath}
\usepackage{amsfonts}
\usepackage{enumitem}
\usepackage{color}
\usepackage{colortbl}
\usepackage{tikz}
\usetikzlibrary{arrows.meta, positioning}
\usepackage{xcolor}
\usepackage{subfigure} 
\usepackage{subcaption}
\usepackage{hyperref}
\hypersetup{                   
  colorlinks,
  linkcolor={green!80!black},
  citecolor={red!70!black},
  urlcolor={blue!70!black}
}
\usepackage{cleveref}
\usepackage{soul}  % for highlighting without wrapping problems of package 'color'
\definecolor{hlgray}{gray}{0.85}
\sethlcolor{hlgray}

\def\BibTeX{{\rm B\kern-.05em{\sc i\kern-.025em b}\kern-.08em
    T\kern-.1667em\lower.7ex\hbox{E}\kern-.125emX}}

\definecolor{codepurple}{rgb}{1,0,1}
\usepackage[most]{tcolorbox}
\usepackage{tikz}

\lstdefinestyle{mystyle}{
  commentstyle=\color{codegreen},
  keywordstyle=\color{magenta},
  stringstyle=\color{codepurple},
  basicstyle=\ttfamily\scriptsize,
  breakatwhitespace=false,
  breaklines=true,
  captionpos=b,
  keepspaces=true,
  showspaces=false,
  showstringspaces=false,
  showtabs=false,
  tabsize=2
}

\usepackage[normalem]{ulem}

\usepackage{xcolor,colortbl}

\usepackage{xargs}
\usepackage[colorinlistoftodos,prependcaption,textsize=normalsize]{todonotes}
\usepackage{cleveref}

\newboolean{COMMENTSON} 
\setboolean{COMMENTSON}{true}   
\ifthenelse{\boolean{COMMENTSON}}
{

}

\usepackage[noend,ruled,linesnumbered]{algorithm2e}
\usepackage{booktabs}
\usepackage{listings}
\usepackage{minted}
\setminted[java]{ %
	linenos=true,             % Line numbers
	autogobble=true,          % Automatically remove common white space
	frame=lines,
	framesep=2mm,
	tabsize=2,obeytabs,
	fontsize=\scriptsize%\footnotesize
}

\usepackage[skip=1pt]{caption}

\newcommand{\tech}{\mbox{{\textsc{CweAgent}}}}

\usepackage{diagbox}
\usepackage{pifont}

\usepackage{multirow}

\usepackage{tabularx}
\usepackage{multirow}
\definecolor{delcolor}{rgb}{1.0, 0.8, 0.8}
\definecolor{addcolor}{rgb}{0.8, 1.0, 0.8}
\definecolor{deltext}{rgb}{0.86, 0.08, 0.24}
\definecolor{addtext}{rgb}{0, 0.5, 0}

\lstdefinelanguage{diff}{
  morecomment=[f][\color{blue}]{@@},
  morecomment=[f][\color{deltext}]{-},
  morecomment=[f][\color{addtext}]{+},
  morecomment=[f][\color{deltext}]{---},
  morecomment=[f][\color{addtext}]{+++},
}

\lstdefinestyle{mystyle}{
    basicstyle=\ttfamily\scriptsize,
    columns=fullflexible,
    backgroundcolor=\color{white},
    numbers=none,
    showstringspaces=false,
    escapeinside={(*@}{@*)},
    frame=none,
    keywordstyle=\color{blue},
    moredelim=[is][\color{deltext}\bfseries\colorbox{delcolor}]{\%-}{-\%},
    moredelim=[is][\color{addtext}\bfseries\colorbox{addcolor}]{\%+}{+\%},
    language=diff,
    aboveskip=0pt,
    belowskip=0pt,
    aboveskip=-3pt,  % Adjust as needed
    belowskip=-3pt,   % Adjust as needed
}

\begin{document}

%\title{Mislabeled at the Source: A Large-Scale Measurement of CWE Labeling Quality in the NVD}
\title{How Reliable Are NVD CWE Labels?\\A Large-Scale Semantic Audit with Seclometry}

\author{%
\IEEEauthorblockN{Yu Nong}
\IEEEauthorblockA{University at Buffalo\\
yunong@buffalo.edu}
\and
\IEEEauthorblockN{Yao Du}
\IEEEauthorblockA{Macau University of Science and Technology\\
yaodu19@gmail.com}
\and
\IEEEauthorblockN{Majid Behravan}
\IEEEauthorblockA{Virginia Tech\\
behravan@vt.edu}
\and
\IEEEauthorblockN{Haipeng Cai}
\IEEEauthorblockA{University at Buffalo\\
haipengc@buffalo.edu}
}

\maketitle
% \affiliation{%
% \affiliation{%

% \affiliation{%
% \affiliation{%


\newcolumntype{P}[1]{>{\raggedright\arraybackslash}p{#1}} % left, wrapped
\newcolumntype{C}[1]{>{\centering\arraybackslash}p{#1}}  % centered, wrapped
\newcolumntype{R}[1]{>{\raggedleft\arraybackslash}p{#1}} % right, wrapped
% Make tag lists break nicely (prevents overlap into next column)
\newcommand{\tagsep}{,\allowbreak\ } % comma + optional line break + space

\begin{abstract}

CWE labels in the National Vulnerability Database (NVD) are widely treated as ground truth for vulnerability search, scanner evaluation, benchmark construction, learning-based security tools, and vulnerability prioritization. Yet their reliability has not been systematically measured at scale, despite growing concerns about NVD’s enrichment backlog and anecdotal reports of inaccurate, ambiguous, or missing labels.
This paper presents a large-scale, code-semantics-grounded measurement of CWE labeling quality in NVD. We build CWEAgent, a validated auditing instrument based on seclometry, a structured representation of vulnerability semantics that captures the root cause, trigger condition, violated security property, exploit mechanism, and impact of vulnerable code. On a manually curated benchmark of 100 open-source CVEs, CWEAgent achieves 85\% top-1 accuracy and 92\% ambiguity-aware accuracy.
Applying CWEAgent to 15,556 open-source CVEs disclosed from 2017–2026, we find that only 49.70\% of NVD CWE labels exactly match the code-grounded label. Another 31.37\% are defensible alternatives under taxonomy ambiguity, while 3.63\% are evidence-inconsistent likely errors. Label reliability varies sharply by assigning organization and weakness type, and apparent project- or language-level differences are largely composition effects of those underlying weakness types. Evidence-inconsistent labels have also increased over time. Through manual review of 434 confirmed mislabels, we identify six recurring error patterns, showing that CWE noise is a structural problem in vulnerability metadata rather than isolated annotation mistakes.

%These findings have direct implications for NVD enrichment, CNA practices, vulnerability benchmarks, and ML-based security research that relies on NVD-derived CWE labels as ground truth.

\end{abstract}

%    <concept>
%        <concept_id>10002978.10003022</concept_id>
%        <concept_desc>Security and privacy~Software and application security</concept_desc>
%        <concept_significance>500</concept_significance>
%        </concept>
%  </ccs2012>
% \ccsdesc[500]{Security and privacy~Software and application security}

% %
% % Keywords. The author(s) should pick words that accurately describe
% % the work being presented. Separate the keywords with commas.

% \keywords{Vulnerability, Patching, Patch Quality, Bad Patches}

\thispagestyle{plain}
\pagestyle{plain}
\pagenumbering{arabic}

\vspace{-1pt}
\section{Introduction}
\vspace{-4pt}
Vulnerability databases are a key security infrastructure, widely used by risk scanners, asset-management systems, threat-intelligence platforms, benchmark builders, and learning-based security tools. Among them, the National Vulnerability Database (NVD)~\cite{nvd} is the dominant public source of enriched CVE metadata. %, assigning structured fields such as affected platforms, severity scores, and Common Weakness Enumeration (CWE) labels~\cite{nvd}. As these fields are machine-readable and widely reused, errors in NVD are not just local documentation mistakes; they propagate broadly into downstream tools, datasets, and decisions.
%
%This paper studies one especially consequential field: the CWE label. 
In particular, a %CWE 
Common Weakness Enumeration (CWE) 
label in NVD identifies the type of weakness underlying a CVE, % It is the vulnerability ecosystem's de facto type system: it determines 
determining how vulnerabilities are grouped, searched, counted, benchmarked, and learned from. Developers use CWE labels to understand root causes and reason about remediations; practitioners use them to organize detection and triage; researchers use them as ground truth when building vulnerability datasets and evaluating weakness-specific techniques. 
% If a CWE label %describes the consequence rather than the cause, confuses sibling weakness types, or assigns a coarse category where a specific weakness is evident, 
% is incorrect, downstream users inherit a distorted view of the vulnerability. 
Due to the widespread use of these labels, noise in them 
%Noise in  %are not just local documentation mistakes; they 
can propagate broadly into downstream tools, datasets, and decisions.

Yet CWE labeling is non-trivial, but a \textit{semantic task}, MITRE's mapping guidance instructs labelers to map a CVE to the most specific applicable root-cause weakness and to avoid labels that merely describe technical impact~\cite{mitre-rcm-guidance}. CVE records, however, are often written in incident language: they describe what an attacker can achieve, which component is affected, or what symptom was observed. However, the code and patch may reveal a different mechanism. For example, a record 
%may describe information exposure while the patch shows cryptographic nonce reuse; it 
may describe a heap buffer overflow while sanitizer output and the patch show an out-of-bounds read. 
%; or it may describe account takeover while the mechanism is improper identity binding. 
Correct labeling therefore requires recovering the weakness mechanism from the advisory, patch, and code, then mapping that mechanism into a taxonomy whose entries are hierarchical and \textit{sometimes overlapping}.

Despite the importance of this task, the reliability of NVD CWE labels has not been systematically measured at scale. Prior work has exposed quality issues in NVD-derived vulnerability data~\cite{croft2023data} and identified administratively invalid CVE--CWE mappings, such as assignments to discouraged or prohibited CWE entries~\cite{simsek2025pocgen}. 
These studies are valuable, but they do not answer the question most relevant to downstream users: \textit{given the available evidence for a CVE, is NVD's assigned CWE semantically consistent with the weakness the evidence supports?} 

Existing automated CWE classifiers are also not sufficient for this measurement. Most predict a CWE from CVE descriptions~\cite{cleaning-nvd} or their diffs~\cite{aota2020automation}, or the structure of CWE catalog itself~\cite{pan2023fine}; they optimize agreement with existing labels \textit{rather than audit those labels}. Training on NVD-labeled data to evaluate NVD label quality creates \textit{a circularity}, and a classifier that only predicts a label cannot tell whether disagreement with NVD reflects \textit{a true error, a defensible alternative, or ambiguity in the CWE taxonomy}.

In this paper, we aim to fill the gap via a large-scale, \textit{code-semantics-grounded} measurement of CWE labeling quality in NVD. 
%Our goal is not to replace NVD analysts or to declare every non-exact match wrong. Instead, we ask: when NVD assigns a CWE to a CVE, how often is that label the code-grounded weakness type, how often is it defensible under taxonomy ambiguity, and how often does it contradict the evidence?
%
%To enable such a measurement at scale, we develop {\tech}, a code-semantic CWE label auditing instrument. % that classifies . 
%It represents both CVEs and CWE entries using \emph{seclometry}, a novel, structured signature of vulnerability semantics covering root cause, trigger condition, violated invariant, source/sink roles, impact, code pattern, and boundaries against neighboring weaknesses. It predicts the CWE most consistent with a CVE's advisory, patch, and code evidence, then separately adjudicates whether the NVD label is exact, defensibly alternative, ambiguous, or evidence-inconsistent. This separation is essential as a mismatch can reflect taxonomy overlap rather than an NVD error.
To enable the study, we %build {\tech}, 
need an automated auditing instrument that classifies vulnerabilities by reasoning over the evidence available for a CVE (e.g., its description, advisory, patch, and relevant code) rather than relying on the description or other metadata alone. The purpose of %{\tech} 
this instrument 
is not to replace NVD analysts or declare every disagreement an error. Instead, it offers a scalable way to compare an NVD-assigned CWE label against the weakness mechanism supported by code-level evidence.
Achieving these, however, faces several key challenges. 

First (\textit{Challenge 1}), the label source cannot also be the training oracle. A natural approach to the instrument is to train a supervised CWE classifier on CVEs labeled by NVD. But this creates a circularity: the same labels whose quality we seek to measure would define the model's notion of correctness. It also risks learning NVD's historical labeling habits, including its coarse labels, description-level shortcuts, and systematic confusions. %{\tech} avoids 
We overcome 
this by not training on NVD-labeled CVEs. Instead, we construct our CWE-side knowledge directly from MITRE's CWE catalog and matches each CVE to that catalog through evidence-grounded vulnerability semantics reasoning at code level.

Second (\textit{Challenge 2}), the relevant semantics are rarely contained in the CVE description alone.
CVE descriptions often describe consequences, exploit effects, or affected components, while the correct CWE depends on the root-cause mechanism. The decisive evidence may appear only in the advisory, fixing patch, sanitizer trace, or vulnerable code. Thus, a description-only classifier may reproduce the same surface cues that led to the original label. %{\tech} addresses 
We address 
this with \emph{seclometry}, a novel, structured security-semantic signature that represents \textit{both CVEs and CWE entries} using the same fields: root cause, trigger condition, violated invariant, source/sink roles, impact, code pattern, and boundaries against neighboring weaknesses. This converts CWE labeling from %free-form 
text matching into %evidence-grounded 
code-semantic alignment.

Third (\textit{Challenge 3}), disagreement is not equivalent to error. 
The CWE taxonomy is hierarchical and partially overlapping: a vulnerability may admit a specific root-cause CWE, a broader parent, or a neighboring CWE that is defensible from another perspective. %A measurement
Thus, our instrument must %therefore 
distinguish imprecision from mislabeling. %{\tech} separates 
We tackle this by \textit{separating
prediction from adjudication}, %. It has 
using 
a \textit{Classifier} to predict the CWE most consistent with the CVE evidence, and an \textit{Arbitrator} to then judge the NVD label as exact, overlap-ambiguous, defensibly alternative, or evidence-inconsistent. This separation lets our measurement quantify both the %large population of 
imprecise-but-defensible labels and the %smaller tail of 
labels that contradict the evidence.

Following these insights, we developed {\tech}, and validate it before using it for measurement. 
%On a manually curated benchmark of 100 open-source CVEs with high-confidence CWE labels, 
On a benchmark of 100 CVEs with manually curated ground-truth CWE labels, 
{\tech} achieves 85\% top-1 exact-match accuracy and 92\% ambiguity-aware accuracy, outperforming same-evidence single-prompt LLM baselines and 
%the supervised diff-based classifier TreeVul~\cite{pan2023fine}. 
the state-of-the-art non-LLM-based classifier~\cite{pan2023fine}. 
We separately evaluate the Arbitrator on 100 adjudication cases containing confirmed-correct and confirmed-mislabeled labels, where it achieves 90\% overall accuracy. These evaluations do not make {\tech} an oracle, but they establish it as a validated measurement instrument; for the final error analysis, we further manually review the likely-mislabel cases.

We then apply this approach to 15,556 open-source CVEs disclosed between 2017 and 2026 with publicly accessible fix commits. Among others, %our noteworthy findings are summarized below.
we revealed that

\begin{itemize}[noitemsep,leftmargin=*,topsep=2pt]
    \item \textit{NVD CWE labels are not simply right or wrong, but fall into exact, defensibly imprecise, and evidence-inconsistent cases.}
    Only {49.70\%} of the CVEs have labels exactly match the code-grounded CWE identified by {\tech}. 
    Yet most disagreements are not outright errors: {25.01\%} reflect overlap ambiguity and {6.36\%} are defensibly alternative labels. %, yielding \textbf{81.07\%} ambiguity-aware correctness. 
    The true error tail is smaller but consequential: {3.63\%} are evidence-inconsistent cases. 
    %likely mislabels. %Thus, treating every non-exact match as wrong overstates NVD error, while treating NVD labels as exact ground truth understates their imprecision.

    \item \textit{CWE label quality has degraded in the error tail over time despite stable exact-match rates.}
    From 2017 to 2026, strict exact-match rates fluctuate without a clear monotonic trend, but evidence-inconsistent labels rise from roughly {1--3\%} in 2017--2018 to {3--6\%} in 2021--2026. %Headline agreement therefore masks a growing tail of labels that contradict the CVE's advisory, patch, or code evidence.
    %The overall exact-match rate looks stable, but the truly wrong labels are becoming more common.

    \item \textit{Weakness semantics, not project or language, primarily explains observed label quality.}
    Exact-match precision varies sharply by CWE family and specific CWE, while evidence inconsistency is flatter at the family level and concentrated in particular CWEs. 
    %Apparent project- and language-level differences are largely composition effects: corpora look better or worse mainly because they contain different weakness mixtures, with injection-heavy corpora appearing more precise and memory-safety-heavy corpora less precise.
    Apparent project- and language-level differences are largely composition effects: a project or language group may appear better or worse labeled mainly because it contains a different mixture of weakness types, with injection-heavy groups appearing more precise and memory-safety-heavy groups less. % precise.

    \item \textit{Assigning CNA is the strongest metadata-level signal of CWE label quality.}
    Label precision varies more across assigning organizations than across years, projects, languages, severity levels, or exploitability indicators. 
    %Because CNA identity is available in CVE metadata, provenance offers an actionable confidence signal for consumers who otherwise treat NVD-derived CWE labels uniformly.
    As each CVE records its %assigning 
    CNA, downstream users can use the CNA as a practical signal of CWE-label reliability instead of assuming all %NVD-derived 
    NVD CWE labels are equally trustworthy.

    \item \textit{Vulnerability importance does not imply label reliability.}
    %CVSS severity does not predict CWE-label quality: 
    Critical-severity CVEs are labeled no more accurately than medium-severity ones. Exploitation status shows the same pattern: CVEs with public Exploit-DB proof-of-concept exploits have an evidence-inconsistency rate statistically indistinguishable from the corpus-wide rate. 
    %Thus, vulnerabilities that drive operational response do not necessarily carry more trustworthy weakness labels.

    \item \textit{%Manual review confirms that 
    The evidence-inconsistent tail contains real NVD mislabels with recurring semantic failure modes.}
    Of the {564} CVEs flagged as evidence-inconsistent, manual review confirms {434} genuine NVD mislabels. Open coding identifies six recurring patterns: \emph{consequence-vs.-root-cause confusion}, \emph{sibling sub-type confusion}, \emph{injection sink/interpreter confusion}, \emph{access-control and identity conflation}, \emph{discouraged or wrong-branch labels}, and \emph{multi-CWE accumulation}. These patterns arise where the taxonomy offers a coarser, easier, or consequence-oriented label than the mechanism shown by the patch or code.

    \item \textit{Confirmed mislabels are patterned but diffuse, making simple correction rules unlikely to suffice.}
    The {434} confirmed mislabels contain {384} distinct directed corrections from the NVD-assigned CWE to the evidence-supported CWE, and {94.8\%} of these correction pairs occur only once. Moreover, {80.5\%} cross CWE-699 weakness families, meaning true mislabels often name a fundamentally different weakness rather than merely a sibling CWE. %This tail calls for evidence-grounded semantic auditing rather than a small hand-written list of CWE swaps.

    \item \textit{Many problematic labels are not evidence-inconsistent, but still impose downstream cost.}
    Over {16\%} of CWE assignments %in our corpus 
    use entries that MITRE discourages or prohibits for direct mapping~\cite{mitre-rcm-guidance}. These labels are rarely evidence-inconsistent, but they are maximally imprecise: they force downstream users to recover the actual weakness type from the advisory, patch, or code instead of the CWE field. We also observe ecosystem-level failure modes:
    \begin{enumerate}[noitemsep,leftmargin=*,topsep=2pt]
        \item CNA and NVD labels %sometimes 
        may disagree on the same CVE;
        \item Later NVD updates can introduce wrong CWE labels even when a correct CWE is already present; and
        \item The original advisory/report may miss the (e.g., root-cause) %root-cause mechanism 
        information needed for correct CWE labeling.
    \end{enumerate}
\end{itemize}

Beyond the measurement itself, we examine a secondary use of {\tech}: assisting CWE-label enrichment under backlog pressure. This analysis is deliberately scoped to the CWE-labeling subtask, not the full NVD enrichment workflow. Using measured per-CVE runtime and monthly CVE publication volume, we simulate how automated candidate labeling could increase CWE-labeling throughput. The result suggests that {\tech} can support human analysts by pre-labeling records, prioritizing review, and flagging evidence-inconsistent assignments, while leaving final enrichment decisions to human review.

In summary, main contributions of this paper include:

\begin{itemize}[noitemsep,leftmargin=*,topsep=2pt]
    \item We formulate \emph{semantic CWE-label auditing} as a measurement problem: determining if an NVD-assigned CWE is exact, defensibly imprecise, or evidence-inconsistent.

    \item We introduce {\tech}, a code-semantics-grounded auditing instrument, and \emph{seclometry}, a structured representation that aligns CVE evidence with MITRE CWE definitions without training on NVD labels.

    \item We validate {\tech}'s Classifier and Arbitrator on controlled benchmarks and then use it to measure CWE-label quality across {15,556} open-source CVEs disclosed between 2017 and 2026.

    \item We characterize the structure and causes of CWE label noise, including {434} manually confirmed NVD mislabels, and derive implications for NVD maintenance, CNA labeling, benchmark construction, and research that treats NVD-derived CWE labels as ground truth.
\end{itemize}

\begin{figure*}[tp]
\centering
\vspace{0pt}
    \includegraphics[width=0.87\linewidth]{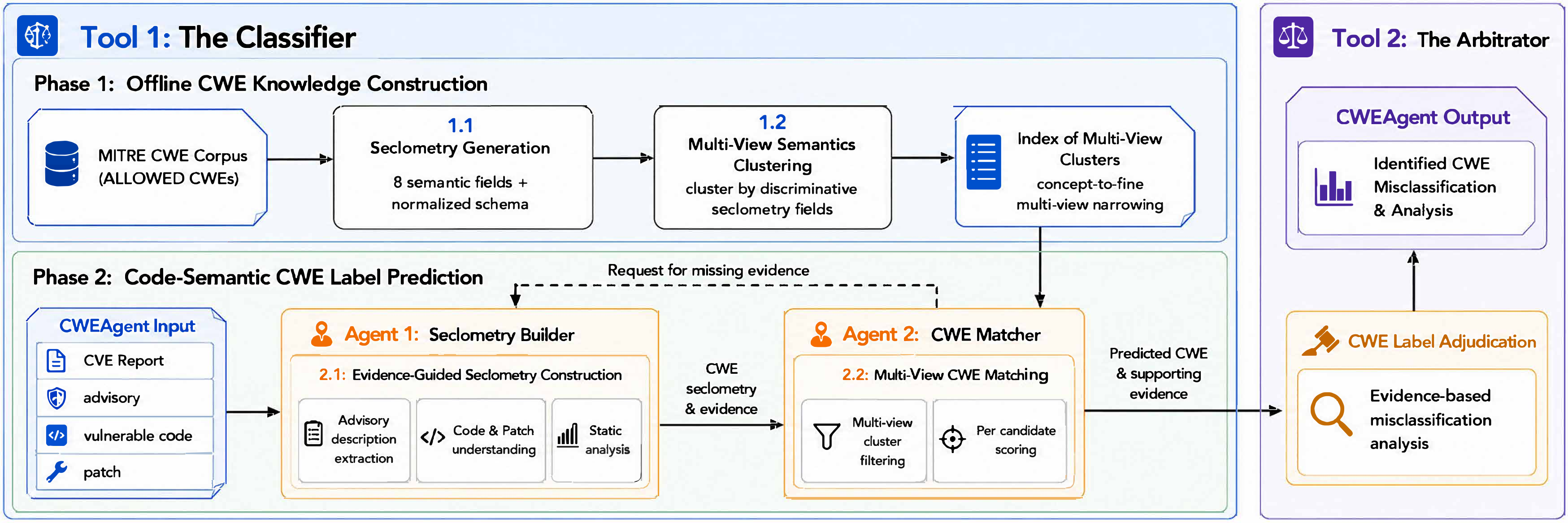}
        \vspace{0pt}
        \caption{An overview of {\tech}'s design, including its inputs, main working phases/steps, and outputs.}
	\label{fig:overview}
        \vspace{-12pt}
\end{figure*}

\vspace{-4pt}
\section{Background and Motivation}
\label{sec:bg}
\vspace{-4pt}
\subsection{CWE-Label Correctness}
\label{sec:bg:correctness}
\vspace{-4pt}
A CWE label is intended to identify the weakness that causes a vulnerability, not merely the symptom described in a CVE record. 
%MITRE's root-cause mapping guidance instructs labelers to choose the most specific applicable CWE, preferring Base or Variant entries when possible and using broader Class-level entries only when no more specific weakness fits~\cite{mitre-rcm-guidance}. 
MITRE’s guidance treats CWE assignment as \textit{root-cause mapping}: labelers should choose the most specific applicable CWE, preferring Base or Variant entries over broader Class-level entries when possible~\cite{mitre-rcm-guidance}.
%The guidance also 
It also marks CWE entries as \textsc{Allowed}, \textsc{Discouraged}, or \textsc{Prohibited} for direct mapping. %; our instrument selects from the \textsc{Allowed} mapping space.

This makes CWE-label correctness a semantic property. A CVE may say that an attacker can disclose data, gain privileges, bypass authentication, or execute code, but those are often consequences. The CWE should capture the mechanism that makes the consequence possible: missing authentication, improper bounds checking, unsafe deserialization, nonce reuse, command injection, or another root-cause weakness. Thus, a label can be plausible from the description yet unsupported by the advisory, patch, or code.

\vspace{-4pt}
\subsection{Motivating Example}
\label{sec:bg:motiv}
\vspace{-4pt}
To illustrate the distinction between consequence and root cause, consider CVE-2026-26014~\cite{cve-2026-26014} (Figure~\ref{fig:case-consequence}) as an example. 
The vulnerability involves AES-GCM encryption with randomly generated nonces. AES-GCM requires nonce uniqueness under a fixed key; a repeated nonce--key pair can enable recovery of the GHASH authentication key and message forgery. The root-cause weakness is therefore \textit{nonce reuse}, corresponding to CWE-323 (Reusing a Nonce, Key Pair in Encryption). The exposure of key material is a consequence of that misuse, not the weakness itself.

The NVD record, however, is labeled CWE-200 (Exposure of Sensitive Information). This label is understandable from a description that emphasizes exposed key material, but it misses the cryptographic mechanism shown by the evidence. 
%
%A benchmark, detector, or measurement that selects CVEs by CWE-323 would therefore miss this case, even though it exhibits precisely that weakness. 
A tool, training benchmark, or prioritization policy that selects vulnerabilities by CWE-323---to study cryptographic
nonce misuse, build a detector for it, or count its prevalence---would miss this CVE. %, because the record does not carry that label.
Conversely, analyses of CWE-200 would incorrectly include a cryptographic nonce-reuse flaw as a generic information-exposure case. 

This example captures the failure mode we study at scale: surface descriptions can point toward broad \textit{consequence} labels, while advisory, patch, and code evidence support more specific \textit{root-cause} CWEs.
%This is the kind of distortion our measurement targets: surface descriptions can point toward broad consequence labels, while advisory, patch, and code evidence support more specific root-cause CWEs.

\begin{figure}[t]
\begin{tcolorbox}[left=3pt,right=3pt,top=2pt,bottom=3pt,
  lifted shadow={1mm}{-2mm}{3mm}{0.1mm}{black!50!white},
  arc=0pt,auto outer arc,boxrule=.5pt,leftrule=2pt,
  title={\textbf{Case 1 (Consequence vs.\ root cause): CVE-2026-26014}},
  fonttitle=\footnotesize, fontupper=\scriptsize]
\vspace{-2pt}
\textbf{NVD label:} CWE-200 (Exposure of Sensitive Information)

\textbf{NVD description.} ``\dots\ uses \hl{random nonce generation}
with AES GCM ciphers, which makes it easier for remote attackers to
obtain the authentication key and spoof data by the
\hl{reuse of a nonce} in a session and a `forbidden attack'.''
\vspace{1pt}
\hrule
\vspace{1pt}
\textbf{Inconsistency.} AES-GCM requires a unique nonce per key; random
nonces admit collisions, and a repeated (nonce, key) pair lets an
attacker recover the GHASH key and forge messages. The root cause is
\hl{nonce reuse}, i.e.\ \textbf{CWE-323 (Reusing a Nonce, Key Pair in
Encryption)}, not the key exposure it enables. NVD's \textbf{CWE-200}
names that downstream \emph{consequence}; root-cause mapping requires the
cryptographic misuse.
\vspace{-4pt}
\end{tcolorbox}
\caption{Consequence-vs-root-cause mislabel: NVD labels the exposure
(CWE-200), not the root cause (CWE-323).}
\vspace{-15pt}
\label{fig:case-consequence}
\end{figure}

\vspace{-4pt}
\subsection{Limitations of Existing Approaches}
\label{sec:bg:whynot}
\vspace{-4pt}
Prior work exposes quality issues in vulnerability metadata~\cite{cleaning-nvd,simsek2025pocgen}, including those in the NVD, but does not measure the \textit{semantic correctness} of assigned CWE labels. 
%NVD-quality studies have shown that CWE fields can be missing or inconsistently populated~\cite{anwar2021cleaning}, and recent work detects administratively invalid CVE--CWE mappings, such as assignments to \textsc{Discouraged} or \textsc{Prohibited} entries~\cite{simsek2025pocgen}. These checks identify important symptoms, but they do not answer whether an assigned CWE is supported by the advisory, patch, and code.
%
%Existing classifiers are also insufficient as auditing instruments for four reasons. 

Meanwhile, existing automated CWE classifiers~\cite{cve2cwe2024,threatzoom,cve2cwe}
advance the state of the art on the \emph{classification} problem but are not sufficient 
%as measurement instruments for four reasons. 
%for our study 
as measurement instruments 
because they fail one or more \ul{requirements} of semantic CWE-label measurement.
%First, description-based classifiers can reproduce the same consequence-oriented cues that led to the original label. 
\textbf{First}, description-based classifiers may help infer missing labels, but they are not suited for auditing assigned labels whose errors may originate from the same description-level cues---the description is authored often by
the same CNA; its vocabulary is entangled with, and predictive of, the existing label. 
%Second, classifiers trained on NVD labels inherit the oracle problem. 
\textbf{Second}, classifiers trained on NVD labels inherit the oracle problem: the labels under audit also define the training target---NVD is the dominant vulnerability data source. 
%Models covering only frequent CWEs cannot audit the long tail of allowed mappings. 
\textbf{Third}, some classifiers are trained and evaluated only on frequent CWE classes, % This may be acceptable for prediction benchmarks, but insufficient 
which fall short 
for auditing NVD---NVD can assign long-tail CWE labels.
%Black-box predictors also do not provide the evidence-grounded explanations needed to distinguish defensible imprecision from true mislabeling
\textbf{Fourth}, black-box predictions cannot distinguish defensible imprecision from evidence-inconsistent mislabeling. 
A measurement audit cannot rely on a label alone. When an instrument disagrees with NVD, it must explain which evidence supports the alternative CWE and \textit{why} the NVD label is exact, defensible, ambiguous, or wrong. 
Prior classifiers generally return predictions without program-evidence-backed rationales, making their disagreements hard to inspect and unsuitable as the basis for a large-scale label-quality measurement.

% Existing classifiers are also insufficient as auditing instruments because they fail one or more requirements of semantic CWE-label measurement.
% First, description-based classifiers can reproduce the same consequence-oriented cues that led to the original label. Second, classifiers trained on NVD labels inherit the oracle problem: the labels under audit also define the training target. Third, restricted-label classifiers cannot audit the long tail of \textsc{Allowed} CWE mappings. Fourth, black-box predictions cannot distinguish defensible imprecision from evidence-inconsistent mislabeling~\cite{v2wbert,threatzoom,cve2cwe}. 

These limitations motivate our {\tech} approach: a code-semantics-grounded CWE label auditing instrument that constructs CWE knowledge independently of NVD labels, reasons over advisory, patch, and code evidence, covers the entire mappable (\textsc{Allowed}) CWE space, and separates (CWE label) prediction from adjudication.

\vspace{-6pt}
\section{Design of {\tech}}
\vspace{-6pt}

\subsection{Overview}
\label{sec:design:overview}
\vspace{-6pt}

Figure~\ref{fig:overview} gives an overview of {\tech}, which consists of two tools. The \emph{Classifier} predicts the CWE most supported by a CVE's report, advisory, patch, and (vulnerable) code evidence as the \textcolor{blue}{\textbf{inputs}}. %, without conditioning on the CVE's existing NVD label. 
The \emph{Arbitrator} then performs \emph{CWE label adjudication}: it compares the existing NVD label against the evidence and the Classifier's prediction, and determines whether the label is exact, overlap-ambiguous, defensibly alternative, or evidence-inconsistent. This separation is central to our measurement: prediction identifies the evidence-supported weakness, while adjudication determines what a disagreement with NVD means.

{\tech}'s design is justified by meeting the requirements of a semantic instrument ($\S$\ref{sec:bg}). 
First, CWE labels must be judged against root-cause semantics rather than surface descriptions. {\tech} therefore uses \emph{seclometry}, a structured representation that expresses both CWE definitions and CVE evidence using the same root-cause and evidence fields (\S\ref{sec:design:seclometry}). Second, the audit must avoid learning from the labels under measurement. {\tech} constructs its CWE-side knowledge offline from MITRE CWE definitions, not from NVD-labeled CVEs (\S\ref{sec:design:phase1}). Third, the audit must reason over evidence rather than accept weak or incomplete inputs. {\tech} separates evidence construction from CWE matching, allowing the matcher to request additional evidence when the current seclometry is insufficient (\S\ref{sec:design:phase2}). Fourth, the audit must distinguish defensible disagreement from true mislabeling. {\tech} therefore separates CWE prediction from label adjudication, using an independent Arbitrator to judge the NVD label rather than treating every mismatch as an error (\S\ref{sec:design:agent3}).

The \ul{Classifier} works in two phases. %In \emph{Phase~1: Offline CWE Knowledge Construction}, 
First (\emph{Phase~1}), it converts the \textsc{Allowed} CWE mapping space into CWE seclometries and organizes them into an \emph{Index of Multi-View Clusters}. This index provides reusable catalog-side structure for coarse-to-fine candidate narrowing. %In \emph{Phase~2: Code-Semantic CWE Label Prediction}, 
Next (\emph{Phase~2}), 
it runs a two-agent workflow for the given CVE. The \emph{Seclometry Builder} %(Agent 1) 
gathers advisory, patch, code, and static-analysis evidence and constructs the CVE seclometry. The \emph{CWE Matcher} %(Agent 2) 
uses the multi-view index to filter candidate CWEs, scores the surviving candidates against the CVE seclometry, and requests missing evidence when needed. %The \textbf{output} is an evidence-supported predicted CWE together with the rationale and evidence used to support it.

%The Arbitrator is the second tool. Given a CVE, its evidence, the Classifier's predicted CWE, and the existing NVD label, it 
Given the Classifier's resulting \textit{predicted CWE \& supporting evidence} and the existing NVD label, 
the \ul{Arbitrator} adjudicates the NVD label under the taxonomy used in our measurement. 
Exact matches are counted as agreement; overlap-ambiguous and defensibly alternative labels are separated from true errors; and evidence-inconsistent labels are flagged for measurement and subsequent manual validation. 

This design lets {\tech} produce not only a CWE prediction, but an auditable judgment about the reliability of the NVD label itself, as the \textcolor{violet}{\textbf{output}} of {\tech}.

\begin{table}[t]
\centering
%\caption{The eight fields of the seclometry's stylometry component and what each is intended to capture.}
\caption{Definition of seclometry: composition \& semantics.}
\label{tab:seclometry-fields}
\setlength{\tabcolsep}{3pt}
\renewcommand{\arraystretch}{0.95}
\footnotesize
\rowcolors{2}{gray!15}{white}
\scalebox{0.85}{
\begin{tabular}{@{}p{0.3\linewidth}p{0.7\linewidth}@{}}
\hline
\textbf{Field} & \textbf{What it captures} \\
\hline
Description & A concise natural-language statement of what the
weakness or vulnerability \emph{is}. \\
Root Cause & The underlying programming or design failure that
allows the weakness to exist. \\
Trigger Conditions & The runtime conditions, including the form of
untrusted input, that must hold for the weakness. \\
Key Invariant Violated & The security property the weakness breaks
(e.g., spatial memory safety, authorization integrity). \\
Observable Effects / Impact & The externally visible consequences
(e.g., DoS, memory corruption, RCE, information disclosure). \\
Typical Code Patterns & Syntactic/idiomatic signatures in source
code that commonly give rise to the weakness. \\
Common Sources / Sinks & The untrusted data sources and the
sensitive sinks characteristic of the weakness. \\
Non-Examples / Boundaries & Explicit distinctions from neighboring
weaknesses that are commonly confused with this one. \\
\hline
\end{tabular}}
\vspace{-15pt}
\end{table}

\vspace{-8pt}
\subsection{Seclometry}
\label{sec:design:seclometry}
\vspace{-8pt}
The central representation in {\tech} is \textbf{seclometry}: a structured security-semantic signature of a CWE definition or a CVE instance. The term is coined by analogy to \emph{stylometry}, which represents documents by measurable stylistic features rather than raw text. Seclometry applies the same idea to vulnerability semantics: it represents a weakness by the security properties that distinguish it from neighboring weaknesses, such as root cause, trigger condition, violated invariant, source/sink role, exploit mechanism, and impact.

%Seclometry is the common coordinate system that makes semantic CWE-label auditing possible. 
Seclometry is the common semantic representation that makes CWE-label auditing possible.
CWE definitions and CVE evidence are written in different genres, and
directly matching these texts encourages surface-level shortcuts on
consequence words such as ``information exposure'' or ``code execution.''
A seclometry instead maps both sides into \textit{the same
security-semantic representation}, letting {\tech} ask whether the
mechanism evidenced by the advisory, patch, and code matches the
mechanism a CWE defines.

\textbf{Design goals.}
A seclometry must satisfy three goals. First, it must be \emph{discriminative}: it should \textit{capture the features that separate} nearby CWEs, not merely broad families. For example, ``memory safety'' is insufficient to distinguish CWE-787 from CWE-125, CWE-416, or CWE-119; the representation must encode whether the flaw is an out-of-bounds write, out-of-bounds read, use-after-free, or broader bounds-management error. Second, it must be \emph{uniform}: the same fields must represent both CWE definitions and CVE evidence, \textit{enabling direct comparison rather than cross-genre text matching}. Third, it must be \emph{machine- and model-consumable}: LLM agents can reason over it, while clustering and matching procedures can compare it systematically.

\textbf{Structure.}
Each seclometry contains two %synchronized 
components. The first is a natural-language security \ul{signature} organized into eight %labeled 
fields (Table~\ref{tab:seclometry-fields}). These fields describe the weakness mechanism, triggering conditions, violated security invariant, observable effects, typical code patterns, source/sink roles, and boundaries against neighboring weaknesses. The second is a normalized \ul{schema} of controlled categorical fields, such as \emph{mechanism\_family}, \emph{source\_type}, \emph{sink\_role}, and \emph{impact\_class}. The %natural-language 
signature supports LLM reasoning and evidence-grounded explanation; the %normalized 
schema supports multi-view clustering, candidate filtering, and systematic scoring. %The two components are redundant by design: each provides a check on the other, and both are present for every CWE and every analyzed CVE.
These fields are derived %from the structure of MITRE CWE entries, including descriptions, common consequences, modes of introduction, observed examples, and relationships, and are 
from the axes along which MITRE's CWE entries are structured (description, common consequences, modes of introduction, observed examples, relationships), and 
informed by the mechanism/operand/consequence decomposition in NIST's Bugs Framework~\cite{nist-bf}. 

We add two fields specifically for CWE label auditing: \emph{Key Invariant Violated} and \emph{Non-Examples/Boundaries}. The invariant field captures the security condition whose violation distinguishes one weakness from another; for instance, CWE-787 and CWE-125 may both involve out-of-bounds memory access, but differ on whether the invalid access writes or reads memory. The boundary field records what a weakness is \emph{not}, allowing {\tech} to penalize candidates that share surface mechanism or impact but contradict the defining boundary of the CWE. This is essential because many NVD disagreements arise %precisely 
at these boundaries.

\begin{figure*}[t]
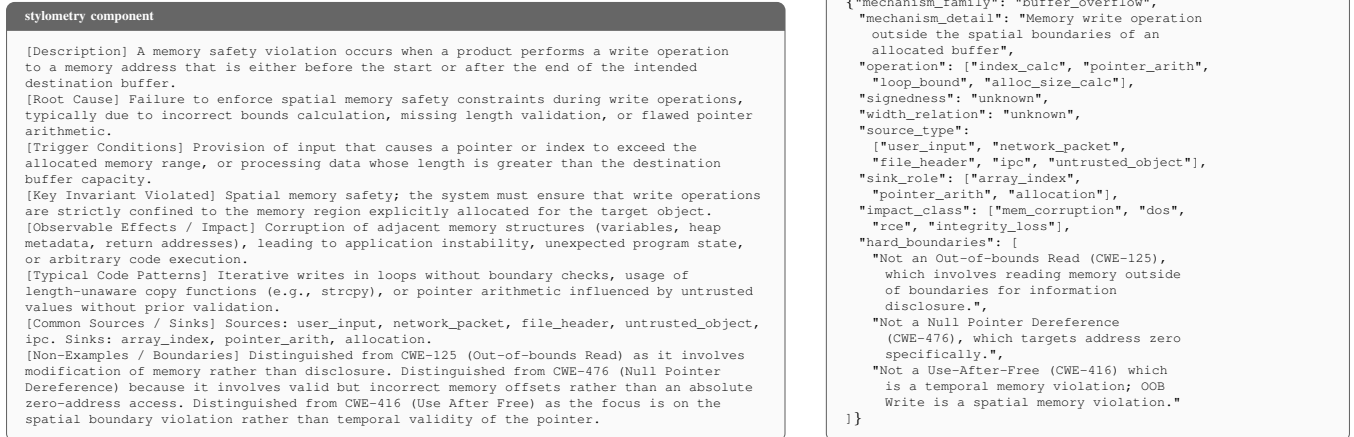

\centering

\begin{minipage}[t]{0.58\linewidth}
\tiny\ttfamily
\begin{tcolorbox}[colback=gray!4,colframe=black!60,boxrule=0.4pt,
left=4pt,right=4pt,top=3pt,bottom=3pt,
title=\textrm{\textbf{stylometry component}}]\selectfont
[Description] A memory safety violation occurs when a product
performs a write operation to a memory address that is either before
the start or after the end of the intended destination buffer.

[Root Cause] Failure to enforce spatial memory safety constraints
during write operations, typically due to incorrect bounds
calculation, missing length validation, or flawed pointer arithmetic.

[Trigger Conditions] Provision of input that causes a pointer or
index to exceed the allocated memory range, or processing data whose
length is greater than the destination buffer capacity.

[Key Invariant Violated] Spatial memory safety; the system must
ensure that write operations are strictly confined to the memory
region explicitly allocated for the target object.

[Observable Effects / Impact] Corruption of adjacent memory
structures (variables, heap metadata, return addresses), leading to
application instability, unexpected program state, or arbitrary
code execution.

[Typical Code Patterns] Iterative writes in loops without boundary
checks, usage of length-unaware copy functions (e.g., strcpy), or
pointer arithmetic influenced by untrusted values without prior
validation.

[Common Sources / Sinks] Sources: user\_input, network\_packet,
file\_header, untrusted\_object, ipc. Sinks: array\_index,
pointer\_arith, allocation.

[Non-Examples / Boundaries] Distinguished from CWE-125
(Out-of-bounds Read) as it involves modification of memory rather
than disclosure. Distinguished from CWE-476 (Null Pointer
Dereference) because it involves valid but incorrect memory offsets
rather than an absolute zero-address access. Distinguished from
CWE-416 (Use After Free) as the focus is on the spatial boundary
violation rather than temporal validity of the pointer.
\end{tcolorbox}
\end{minipage}
\hfill
\begin{minipage}[t]{0.39\linewidth}
\tiny\ttfamily
\begin{tcolorbox}[colback=gray!4,colframe=black!60,boxrule=0.4pt,
left=4pt,right=4pt,top=3pt,bottom=3pt,
title=\textrm{\textbf{normalized schema}}]
\{\hspace*{0em}"mechanism\_family": "buffer\_overflow",\\
\hspace*{1em}"mechanism\_detail": "Memory write operation\\
\hspace*{2em}outside the spatial boundaries of an\\
\hspace*{2em}allocated buffer",\\
\hspace*{1em}"operation": ["index\_calc", "pointer\_arith",\\
\hspace*{2em}"loop\_bound", "alloc\_size\_calc"],\\
\hspace*{1em}"signedness": "unknown",\\
\hspace*{1em}"width\_relation": "unknown",\\
\hspace*{1em}"source\_type": \\ 
\hspace*{2em}["user\_input", "network\_packet",\\
\hspace*{2em}"file\_header", "ipc", "untrusted\_object"],\\
\hspace*{1em}"sink\_role": ["array\_index",\\
\hspace*{2em}"pointer\_arith", "allocation"],\\
\hspace*{1em}"impact\_class": ["mem\_corruption", "dos",\\
\hspace*{2em}"rce", "integrity\_loss"],\\
\hspace*{1em}"hard\_boundaries": [\\
\hspace*{2em}"Not an Out-of-bounds Read (CWE-125),\\
\hspace*{3em}which involves reading memory outside\\
\hspace*{3em}of boundaries for information\\
\hspace*{3em}disclosure.",\\
\hspace*{2em}"Not a Null Pointer Dereference\\
\hspace*{3em}(CWE-476), which targets address zero\\
\hspace*{3em}specifically.",\\
\hspace*{2em}"Not a Use-After-Free (CWE-416) which\\
\hspace*{3em}is a temporal memory violation; OOB\\
\hspace*{3em}Write is a spatial memory violation."\\
\hspace*{0em}]\}
\end{tcolorbox}
\end{minipage}
\caption{Seclometry of \texttt{CWE-787} (Out-of-bounds~Write),
showing both the eight-field stylometry component (left) and the
normalized schema (right). The stylometry component is the
natural-language signature read by LLM-based matching stages; the
normalized schema is used by multi-view clustering and
per-candidate scoring. The same two-component structure applies to
CVE seclometries.}
\label{fig:seclometry-example}
\end{figure*}

%In Phase~1, {\tech} constructs seclometries for all 842 \textsc{Allowed} CWE entries and organizes them into a reusable multi-view index. In Phase~2, Agent~1 constructs the same representation for each CVE from advisory, patch, code, and static-analysis evidence. Agent~2 then compares CVE and CWE seclometries to identify the evidence-supported CWE, and the Arbitrator uses the same representation to decide whether the existing NVD label is exact, defensibly imprecise, ambiguous, or evidence-inconsistent. 
Figure~\ref{fig:seclometry-example} shows a complete seclometry for \texttt{CWE-787} (Out-of-bounds Write) in both components. The left panel is the eight-field stylometry component---the natural-language signature that the LLM-based matching stages read---while the right panel is the normalized schema of categorical fields that multi-view clustering and per-candidate scoring consume as discrete values. The same two-component structure applies to every one of the 842 ALLOWED CWEs and to each CVE seclometry that Agent~1 emits.

\vspace{-6pt}
\subsection{Tool 1: The Classifier}
\label{sec:design:classifier}
\vspace{-6pt}
The Classifier aims to identify the CWE most supported by a CVE's advisory, patch, code, and static-analysis evidence, without conditioning on the CVE's existing NVD label. This prediction is not itself the final label-quality measurement: rather, it provides the evidence-supported candidate CWE label that the Arbitrator later uses to judge the NVD label. 
It works in two phases as elaborated below. 
%Phase~1 is an offline catalog-construction phase that builds a reusable semantic index over MITRE's \textsc{Allowed} CWE mapping space. Phase~2 is a per-CVE prediction phase that maps an incoming CVE into the same seclometry representation and matches it against the indexed CWE catalog.
\vspace{-6pt}
\subsubsection{Phase 1: Offline CWE Knowledge Construction}
\label{sec:design:phase1}

%Phase~1 constructs the catalog-side knowledge used by the Classifier. Its input is the MITRE CWE corpus filtered to the \textsc{Allowed} mapping space, and its output is an \emph{Index of Multi-View Clusters}: a reusable semantic index over all 842 candidate CWEs. 
Phase~1 takes the MITRE CWE corpus filtered to the \textsc{Allowed} mapping space as input and 
constructs the catalog-side knowledge used by the Classifier. 
This phase is offline as the CWE catalog is stable relative to per-CVE inference, and that catalog-wide reasoning should be performed once, inspected, and reused rather than recomputed for every CVE. %Conceptually, Phase~1 turns the CWE catalog from a collection of heterogeneous (XML) entries into a structured, searchable space of weakness (CWE) semantics.

%Phase~1 has two steps. Step~1.1 converts each CWE entry into a seclometry. Step~1.2 clusters the resulting seclometries independently along each semantic view, producing the multi-view index consumed by Phase~2.

% \subsection{The Classifier: Phase 1 --- Offline Knowledge Construction}
% \label{sec:design:phase1}
% Phase~1 is executed once against the MITRE CWE corpus and produces the \emph{Index of Multi-view Clusters} that Phase~2's Matcher consumes. The phase is offline for two reasons. First, the CWE catalog does not change on a per-query basis; recomputing a catalog-wide representation for every incoming CVE would be pure waste. Second, because Phase~1 runs only once, we are free to use the strongest LLM available regardless of per-call cost, whereas Phase~2 must run on every CVE and therefore favors a cheaper model (\S\ref{sec:design:implementation}). Moving the catalog-wide reasoning offline lets us buy quality where it matters most. The phase consists of two steps, \emph{seclometry generation} and \emph{multi-view semantic clustering}, both implemented with Claude~4.6 Opus at temperature~0 for determinism given fixed inputs.

\textbf{Step \textcolor{blue}{\textbf{1.1}}: Seclometry Generation.}
For each \textsc{Allowed} CWE, {\tech} constructs a seclometry containing both a natural-language security signature and a normalized categorical schema. The input is the raw MITRE XML entry, including its description, common consequences, modes of introduction, observed examples, and relationship metadata. These fields provide the raw material for most seclometry fields: the LLM's role is to organize and distill the CWE's mechanism, trigger conditions, violated invariant, code patterns, source/sink roles, impact, and boundaries into the common representation (as defined in $\S$\ref{sec:design:seclometry}).

The most important synthesis occurs in the boundary fields. MITRE entries often describe what a weakness is, but they rarely state explicitly what it is \emph{not}. For auditing, this negative space is crucial: many CWE confusions occur between nearby weaknesses that share a broad mechanism or consequence but differ in the defining invariant. For example, CWE-787 and CWE-125 both involve out-of-bounds memory access, but differ on write versus read. {\tech} therefore derives \emph{Non-Examples/Boundaries} from CWE relationships (\emph{PeerOf}, \emph{CanPrecede}, \emph{CanFollow}), neighboring entries, and common confusion patterns, so that later matching can penalize candidates that fit surface symptoms but violate the CWE's defining boundary.

The output is a collection of 842 CWE seclometries, one per \textsc{Allowed} \textit{CWE}. Each seclometry uses the same two-component format that Agent~1 will later produce for \textit{CVEs}. This uniformity is essential: Phase~2 does not compare CVE descriptions directly to CWE prose, but compares CVE and CWE seclometries in the same semantic representation.

% \tech\ first converts each CWE in the MITRE catalog into the seclometry form. The input is the raw CWE XML filtered to the 842 entries whose MITRE mapping status is \emph{ALLOWED}, i.e., the entries MITRE intends to be used as classification labels. For each such CWE, the raw XML entry is passed to the LLM with a fixed prompt that produces both the eight-field stylometry and the normalized schema in a single structured response. This single-pass formulation works because the CWE XML already contains the raw material for most fields---\emph{Description}, \emph{Common\_Consequences}, \emph{Modes\_of\_Introduction}, and \emph{Observed\_Examples} populate the corresponding seclometry fields almost directly---so the LLM's primary work is organization and distillation rather than invention. The one field that requires genuine synthesis is \emph{Non-Examples~/~Boundaries}, which the LLM derives from the CWE's \emph{Relationships} entries (\emph{PeerOf}, \emph{CanPrecede}, \emph{CanFollow}) together with its own knowledge of which neighboring CWEs are most commonly confused with the target. We apply a lightweight post-processing step to normalize malformed outputs (e.g., JSON parsing fixes) before storage.

% The output of Step~1.1 is a collection of 842 seclometries, one per ALLOWED CWE, each following the two-component structure illustrated in Figure~\ref{fig:seclometry-example}. This collection is the sole input to Step~1.2; the raw MITRE XML is not consulted again.

\textbf{Step \textcolor{blue}{\textbf{1.2}}: Multi-View Semantics Clustering.}
The 842-CWE space is too large and semantically uneven for direct all-against-all reasoning at every CVE. At the same time, a single global clustering would be too coarse: two CWEs may agree on one semantic facet, such as impact, while differing on another, such as violated invariant or source/sink role. {\tech} therefore clusters the catalog \emph{by view}. Each seclometry field defines one \textit{view} of the CWE space, and the CWEs are partitioned separately under that view.

This multi-view design preserves distinctions that a joint embedding or single partition would blur. A candidate may be close to a CVE under the impact view but far under the invariant view; another may share the mechanism but differ in the sink role. Keeping these views separate lets the Matcher perform \textit{coarse-to-fine narrowing by combining evidence across facets}, rather than relying on a single similarity score that averages away the reason for disagreement.

\begin{figure}[t]
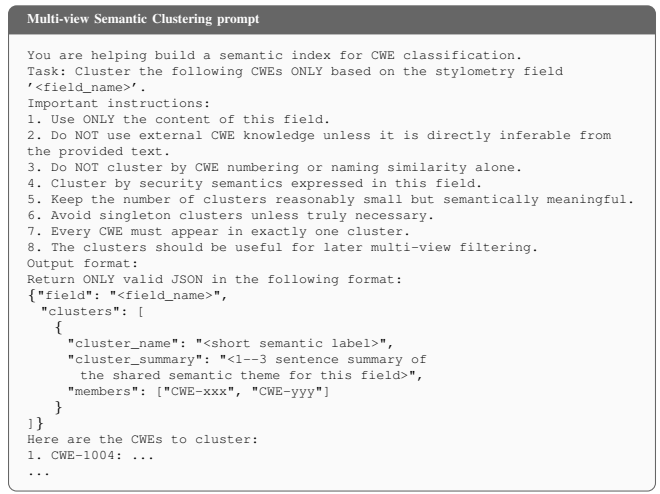

\tiny\ttfamily
\begin{tcolorbox}[colback=gray!4,colframe=black!60,boxrule=0.4pt,
left=4pt,right=4pt,top=3pt,bottom=3pt,
title=\textrm{\textbf{Multi-view Semantic Clustering prompt}}]
\selectfont
You are helping build a semantic index for CWE classification.

Task: Cluster the following CWEs ONLY based on the stylometry
field '<field\_name>'.

Important instructions:\\
1. Use ONLY the content of this field.\\
2. Do NOT use external CWE knowledge unless it is directly
inferable from the provided text.\\
3. Do NOT cluster by CWE numbering or naming similarity alone.\\
4. Cluster by security semantics expressed in this field.\\
5. Keep the number of clusters reasonably small but semantically
meaningful.\\
6. Avoid singleton clusters unless truly necessary.\\
7. Every CWE must appear in exactly one cluster.\\
8. The clusters should be useful for later multi-view filtering.

Output format:\\
Return ONLY valid JSON in the following format:

\{\hspace*{0em}"field": "<field\_name>",\\
\hspace*{1em}"clusters": [\\
\hspace*{2em}\{\\
\hspace*{3em}"cluster\_name": "<short semantic label>",\\
\hspace*{3em}"cluster\_summary": "<1--3 sentence summary of\\
\hspace*{4em}the shared semantic theme for this field>",\\
\hspace*{3em}"members": ["CWE-xxx", "CWE-yyy"]\\
\hspace*{2em}\}\\
\hspace*{0em}]\}

Here are the CWEs to cluster:\\
1. CWE-1004: ...\\
\ldots
\end{tcolorbox}
\caption{Prompt used in Step~1.2 to cluster the 842 ALLOWED CWEs
along one seclometry field. The prompt is instantiated once per
field, eight times in total, with \texttt{<field\_name>} and the
per-CWE field values filled in. The prompt has been lightly edited
from the implementation version for terminology alignment with the
paper; the unedited version is available in the artifact.}
\label{fig:clustering-prompt}
\end{figure}

For each view, the clustering procedure partitions all 842 CWE seclometries into semantically coherent clusters. Each cluster has a short label, a concise theme summary, and a list of CWE IDs. The prompt (Figure~\ref{fig:clustering-prompt})
enforces three constraints: clustering must use only the selected seclometry field, not superficial CWE names or numeric IDs; every CWE must appear in exactly one cluster for that view; and clusters should be semantically meaningful rather than mechanically balanced. This produces field-specific partitions that can later be intersected, ranked, or relaxed during matching (by the CWE Matcher in \textbf{Step~\textcolor{orange}{2.2}}).

This prompt clusters the 842 ALLOWED CWEs along a single seclometry field; it is instantiated once per field, eight times in total. Figure~\ref{fig:clustering-prompt} shows the template. A single-pass call suffices because all 842 per-field values fit within the model's context window; hierarchical or iterative schemes would require an explicit similarity metric---precisely the judgment we delegate to the LLM---and would compound non-determinism across rounds. Temperature is fixed at~0 with fixed prompt and model versions, so each per-field clustering is deterministic given its input.

The output of Phase~1 is the \emph{Index of Multi-View Clusters}. For each seclometry field, the Index records the view-specific clusters and the CWE membership of each cluster; equivalently, for each CWE, it records that CWE's cluster assignment under every view. Phase~2 consumes this Index to narrow the candidate space before per-candidate scoring. Because the Index is built once and reused across all CVEs, it provides catalog-wide semantic structure without adding per-CVE catalog-construction cost.

\vspace{-6pt}
\subsubsection{Phase 2: Code-Semantic CWE Label Prediction}
\label{sec:design:phase2}
For a given CVE under auditing, Phase~2 invokes the Classifier against the CVE record and available references as inputs, and produces the predicted CWE label %, confidence, match-quality label, 
and supporting evidence as outputs. Algorithm~\ref{alg:phase2} details the Phase~2 control flow for a single CVE. Agent~1 builds the CVE seclometry and evidence bundle (line~3); Stage~A filters the 842 candidates to a short list against the Index, and Stage~B scores each survivor (lines~4--6). If the top candidate's confidence is not low, the prediction is returned immediately (lines~7--8). Otherwise the self-reflection step produces targeted evidence queries (line~9) that re-invoke Agent~1 on the next iteration; the loop runs at most three times, an empirically chosen budget beyond which the predicted CWE rarely changes.

\begin{algorithm}[t]
\footnotesize
\setlength{\algomargin}{1.5em}
\caption{Phase~2 (Code-Semantic CWE Label Prediction)}
\label{alg:phase2}
\DontPrintSemicolon
\SetKwInOut{Input}{Input}
\SetKwInOut{Output}{Output}
\Input{CVE record $r$; Index of Multi-view Clusters $I$}
\Output{Predicted CWE $c^\star$, confidence, evidence}
\SetKwFunction{AgentOne}{Agent1}
\SetKwFunction{Filter}{MultiViewFilter}
\SetKwFunction{Score}{PerCandidateScore}
\SetKwFunction{Rank}{Rank}
\SetKwFunction{Reflect}{SelfReflect}
$Q \gets \emptyset$ \tcp*{targeted evidence queries}
\For{$i \gets 1$ \KwTo $3$}{
  $(s_{\text{cve}}, E) \gets \AgentOne(r, Q)$ \tcp*{Agent 1}
  $\mathcal{S} \gets \Filter(s_{\text{cve}}, I)$ \tcp*{Stage A}
  \ForEach{$c \in \mathcal{S}$}{
    $\sigma_c \gets \Score(s_{\text{cve}}, c, E)$ \tcp*{Stage B}
  }
  $(c^\star, \mathit{conf}) \gets \Rank(\mathcal{S}, \{\sigma_c\})$\;
  \If{$\mathit{conf} \ne \text{\upshape low}$}{
    \Return $(c^\star, \mathit{conf}, E)$
  }
  $Q \gets \Reflect(\mathcal{S}, \{\sigma_c\}, s_{\text{cve}}, E)$\;
  \lIf{$Q = \emptyset$}{\Return $(c^\star, \mathit{conf}, E)$}
}
\Return $(c^\star, \mathit{conf}, E)$
\end{algorithm} 
%Whereas Phase~1 builds the catalog-side representation once, Phase~2 constructs the CVE-side representation on demand and matches it against the indexed CWE space. 
%
%Phase~2 uses two agents. Agent~1, the \emph{Seclometry Builder}, gathers evidence and converts the CVE into the same seclometry representation used for CWEs. Agent~2, the \emph{CWE Matcher}, uses the Phase~1 index to narrow the candidate space and then scores the surviving CWEs against the CVE seclometry. When the evidence is insufficient to distinguish close candidates, Agent~2 sends targeted evidence requests back to Agent~1.
%
This phase works in two steps below, each realized by an LLM-based agent. 

\textbf{Step~\textcolor{orange}{2.1}: Evidence-Guided Seclometry Construction (\textcolor{orange}{Agent 1:} Seclometry Builder).}\label{sec:design:agent1}
%
%In this step, 
Agent 1 constructs the seclometry of the given CVE. 
%with evidence anchors recording where each element came from. %Its input---the CVE description, advisory and patch URLs, and any referenced source artifacts---is often incomplete, so the Builder gathers additional evidence when needed rather than relying only on the CVE description.
%
To ground the construction in evidence (e.g., patch URLs and any referenced source artifacts), which is often incomplete, 
the agent uses four tools: \emph{web search} to find advisories when the CVE reference list is thin; a \emph{webpage reader} that follows Wayback snapshots where available to reduce post-disclosure leakage; a \emph{source downloader} for files named in the patch or advisory; and a \emph{CodeQL scanner} that surfaces candidate weakness patterns. CodeQL findings are auxiliary rather than authoritative: a positive finding can support a mechanism, but a null result is not treated as negative evidence.

Agent~1 applies these tools across the three sub-steps shown in Figure~\ref{fig:overview}. \emph{Advisory description extraction} drafts an initial weakness signature from the CVE description and advisories. \emph{Code and patch understanding} locates the root-cause construct in the patch and source. \emph{Static analysis} folds confirmed CodeQL findings into the evidence bundle when they help identify or disambiguate the weakness mechanism.

Agent~1 does not invoke every tool for every CVE. When the available evidence already determines the weakness class unambiguously, it emits the seclometry directly; otherwise, it proceeds through additional evidence-gathering steps. When re-invoked by the refinement loop, Agent~1 receives targeted evidence queries from Agent~2 and focuses its tool use on the distinctions that matter for the remaining candidate CWEs. The Builder emits the CVE seclometry together with evidence anchors, including advisory URLs, file paths, code line ranges, patch excerpts, and CodeQL snippets supporting each claim.

\textbf{Step~\textcolor{orange}{2.2}: Multi-View CWE Matching (\textcolor{orange}{Agent 2:} CWE Matcher).}\label{sec:design:agent2}
This step/agent turns the CVE seclometry into a CWE prediction. It works in two stages. 
%Stage~A performs \emph{multi-view cluster filtering}, using the Index from Phase~1 to narrow the 842-CWE catalog to a short list of plausible candidates. Stage~B performs \emph{per-candidate scoring}: it compares each surviving candidate CWE against the CVE seclometry and selects the evidence-supported prediction. 

\emph{Stage A: multi-view cluster filtering.}
The goal of Stage~A is to reduce the candidate space cheaply while preserving semantically plausible alternatives. For each seclometry field, the Matcher takes the CVE's value for that field and compares it against the cluster labels, summaries, and representative members from Phase~1's partitioning of the catalog under the same view. It then selects the best-matching cluster and, when appropriate, a runner-up cluster. This is \emph{cluster selection}, not re-clustering: the Phase~1 partitions are fixed, and Stage~A only chooses among them.

As the Matcher processes the seclometry fields, each selected cluster contributes to a cumulative \emph{multi-view prior} over CWEs. Candidate CWEs receive support when they appear in clusters selected under multiple views, and the candidate set is narrowed by combining best- and runner-up-cluster memberships across fields. This produces a short list, typically small enough for Stage~B to score each candidate independently. The scoring weights are fixed hyperparameters, not tuned per dataset.

\emph{Stage B: per-candidate scoring.}
Stage~B scores each surviving CWE against the CVE independently. The scoring prompt (Figure~\ref{fig:pairwise-prompt})
receives the CVE seclometry, the candidate CWE seclometry, and, when available, relevant source or patch snippets collected by Agent~1. It returns a compact structured judgment: a numeric score (0--100), confidence level, match-quality label (strong/partial/weak/contradiction), and short justification.

The scoring rules follow directly from the discriminative role of seclometry. The Matcher is instructed to (1) score the primary weakness mechanism rather than downstream consequence; (2) prefer the most specific correct root-cause CWE; (3) penalize contradictions with the candidate CWE's \emph{Non-Examples/Boundaries}; and (4) use source or patch evidence when it helps distinguish neighboring weaknesses. Thus, Stage~B does not merely ask whether a CWE name sounds plausible: it asks whether the candidate's defining mechanism and boundary conditions are supported by the CVE evidence. 
The model reasons step by step internally but outputs only the compact 
judgment---a form of \emph{internal zero-shot
chain-of-thought}~\cite{kojima2022large}---keeping per-candidate cost
low across the short list. 

The prompt's four rules operationalize the discriminative design of the seclometry: score the primary weakness mechanism rather than downstream consequence, prefer the most specific correct root-cause CWE, penalize contradictions with the candidate's Non-Examples/Boundaries, and use source code only when it helps distinguish mechanism. The model reasons step by step internally but emits only the compact structured judgment.

\begin{figure}[h]
\tiny\ttfamily
\begin{tcolorbox}[colback=gray!4,colframe=black!60,boxrule=0.4pt,
left=4pt,right=4pt,top=3pt,bottom=3pt,
title=\textrm{\textbf{Per-candidate scoring prompt}}]
\selectfont
Return JSON only. No extra text.

Task:\\
Score whether the vulnerability matches this SINGLE candidate CWE.

Rules:\\
1) Score the PRIMARY weakness mechanism (root cause), not
downstream consequence.\\
2) Prefer the most specific correct root-cause CWE.\\
3) Penalize contradictions with the candidate boundaries /
non-examples.\\
4) Use source code only if it helps.\\
5) Keep the decision compact.\\
6) Reason step by step internally before answering, but DO NOT
output your private chain-of-thought.

Output JSON:\\
\{\\
\hspace*{1em}"cwe\_id": "<cwe\_id>",\\
\hspace*{1em}"score": <integer 0--100>,\\
\hspace*{1em}"confidence": "low$|$medium$|$high",\\
\hspace*{1em}"match\_type": "strong$|$partial$|$weak$|$contradiction",\\
\hspace*{1em}"note": "<max 20 words>"\\
\}

Vulnerability normalized schema:\\
<<<VNORM\\
\ldots\\
VNORM>>>

Vulnerability stylometry:\\
<<<VSTY\\
\ldots\\
VSTY>>>

Relevant source evidence (when available):\\
<<<SOURCE\\
file\_path: \ldots\\
start\_line, end\_line: \ldots\\
<code snippet>\\
SOURCE>>>

Candidate CWE:\\
ID: <cwe\_id>, Name: <cwe\_name>

Candidate normalized schema:\\
<<<CNORM\\
\ldots\\
CNORM>>>

Candidate stylometry:\\
<<<CSTY\\
\ldots\\
CSTY>>>
\end{tcolorbox}
\caption{Prompt used by Agent~2 to score a single shortlisted CWE
against the CVE. The prompt is instantiated once per candidate on
the Stage~A short list and returns a compact structured judgment.
An internal chain-of-thought instruction
(Rule~6) asks the model to reason step by
step before producing the compact output. The prompt has been
lightly reflowed for the figure; the production version is
available in the artifact.}
\label{fig:pairwise-prompt}
\end{figure}

Each candidate's final score combines its Stage~B match score with the Stage~A multi-view prior. The highest-scoring candidate becomes the predicted CWE. When the top candidate has low confidence or a small margin over the runner-up, the Matcher performs a self-reflection step that generates targeted evidence queries %, such as whether a patch modifies an allocation-size computation or only a bounds check. 
(e.g., ``does the patch modify the allocation-size computation, or only the bounds check?''). 
These queries drive the evidence-refinement loop.

\textbf{Evidence-Refinement Loop.}\label{sec:design:feedback}
This loop lets Agent~2 request additional evidence from Agent~1 when the current evidence cannot distinguish close candidates. The loop triggers when the top-candidate confidence is low or %the margin over the runner-up is small, 
Agent~2's self-reflection produces at least one query. 
%and Agent~2 can articulate a concrete evidence query. 
Agent~1 is then re-invoked with those targeted queries, gathers the requested evidence, and Agent~2 reruns filtering and scoring.
The loop terminates when the prediction reaches medium or high confidence, when Agent~2 produces no new evidence query, or after a fixed iteration cap. The cap prevents ambiguous CVEs from consuming unbounded tool calls. %; in our implementation, we use three iterations. 

\textbf{Scoring weights.} The Matcher combines evidence across the eight views with a small set of fixed weights. In Stage~A, every CWE in the best-matching cluster of a field gains a confidence-weighted contribution to its multi-view prior, and members of the runner-up cluster contribute at a reduced weight; the candidate set is hard-narrowed after each field by intersecting it with the union of the best- and runner-up-cluster members, so a CWE survives only if it appears in one of the two selected clusters for every field. In Stage~B, each candidate's final matching score is a linear combination of its per-candidate score and its Stage~A multi-view prior. All weights are set once and held fixed across every experiment in the paper; we did not tune them per dataset.

\vspace{-6pt}
\subsection{Tool 2: Arbitrator}\label{sec:design:agent3}
\vspace{-8pt}
The Classifier predicts the CWE most supported by the collected evidence, but a prediction-label mismatch is not automatically an NVD error. CWE entries overlap, differ in granularity, and sometimes describe different defensible views of the same vulnerability. Conversely, a mismatch may also reflect a real NVD mislabel rather than a classifier mistake. {\tech}'s second tool, the \emph{Arbitrator}, resolves this ambiguity by judging the existing NVD label against the CVE evidence and the Classifier's prediction.

\textcolor{orange}{\textbf{CWE Label Adjudication.}}
%Given a CVE, the evidence bundle, the Classifier's predicted CWE, and the existing NVD label, the Arbitrator assigns the NVD label to one of four measurement categories:
%
Given a CVE, the evidence bundle, the Classifier's predicted CWE, and the existing NVD label, the Arbitrator judges the NVD label under the measurement taxonomy we use in this paper: exact, overlap-ambiguous, defensibly alternative, or evidence-inconsistent. 
%It then emits one of three operational labels used by our evaluation and measurement pipeline: \emph{count\_as\_correct}, \emph{discard}, or \emph{keep\_as\_incorrect}.

\begin{enumerate}[noitemsep,leftmargin=*,topsep=2pt]
    \item \emph{Exact match.} The NVD label and the predicted, evidence-supported CWE label are the same (\textbf{S1}). 
    %The Arbitrator outputs \emph{count\_as\_correct}.

    \item \emph{Overlap ambiguity.} The NVD label and the predicted CWE overlap for this CVE; both are materially defensible, typically because one is broader, narrower, or adjacent to the other in the CWE taxonomy (\textbf{S2}). 
    %The Arbitrator outputs \emph{count\_as\_correct}.

    \item \emph{Defensibly alternative labeling.} The NVD label is not an exact or overlapping match, but remains supported by the evidence under a plausible interpretation. 
    %such as capturing a higher-level manifestation, a lower-level mechanism, or a closely related root-cause view. 
    This covers cases in which one CWE label captures the root cause while the other captures the resulting weakness or impact, or one captures a lower-level mechanism while the other captures a higher-level manifestation (\textbf{S3}). 
    %The Arbitrator outputs \emph{count\_as\_correct}.

    \item \emph{Evidence-inconsistent labeling.} The evidence supports the predicted CWE while contradicting, or providing no defensible support for, the NVD label (\textbf{S4}). 
    %The Arbitrator outputs \emph{discard}, marking the case as unreliable as a classifier-evaluation reference and as a candidate NVD mislabel for measurement.
\end{enumerate}

The Arbitrator further maps these four \ul{\textbf{S}}ituations into three output labels by a fixed priority rule. If the NVD label is exact, overlap-ambiguous, or defensibly alternative, the output is \emph{count as correct}. If the NVD label is evidence-inconsistent, the output is \emph{discard}. Otherwise, the output is \emph{classifier error}. 
These labels serve distinct downstream roles. \emph{count as correct} supports ambiguity-aware accuracy by treating taxonomy ambiguity and defensible imprecision as non-errors. \emph{discard} removes unreliable reference labels from the classifier-evaluation denominator and marks them as candidate NVD mislabels for the wild-scale measurement. \emph{keep as incorrect} preserves cases where the NVD label is supported and the Classifier prediction is a genuine error.
The adjudication prompt is shown in Figure~\ref{fig:agent3-prompt}. 

This prompt adjudicates a single (CVE, predicted CWE, NVD label) tuple into the Arbitrator's three-way decision. It checks the three situations in priority order---overlap ambiguity, alternative-but-defensible labeling, and reference-CWE inconsistency---and emits the corresponding decision (\texttt{count\_as\_correct}, \texttt{discard}, or \texttt{keep\_as\_incorrect}) over the evidence bundle gathered by the Classifier, without re-deriving evidence of its own.

\begin{figure}[h]
\tiny\ttfamily
\begin{tcolorbox}[colback=gray!4,colframe=black!60,boxrule=0.4pt,
left=4pt,right=4pt,top=3pt,bottom=3pt,
title=\textrm{\textbf{Arbitrator prompt}}]
\selectfont
You are reviewing a CWE classification mismatch for one CVE.
Your job is NOT to predict the best CWE from scratch. Your job
is only to determine whether this mismatch should still count
against the technique.

CVE ID: <cve\_id>\\
Assigned CWE: <ref>\\
Predicted CWE: <pred>

Please check the following three situations.

\textbf{Situation 1: overlap / ambiguity.} Check whether this CVE
genuinely lies in the overlap between the predicted CWE and the
reference CWE, including cases where both CWEs are materially
defensible for this specific vulnerability even if one is more
general and the other more specific. Mark true when the two
substantially overlap for this CVE.

\textbf{Situation 2: alternative-but-defensible labeling.} Check
whether the predicted CWE is clearly supported by the CVE's
description, patch, code, or reasoning evidence, even when it does
not strongly overlap with the reference CWE. This includes
cases where one CWE captures root cause while the other captures
resulting weakness or impact, or where one captures a lower-level
mechanism while the other captures a higher-level manifestation.
Do not require the labeled CWE to be wrong; mark true if the
predicted CWE is clearly evidenced and reasonable.

\textbf{Situation 3: reference CWE inconsistency.} Check whether
there is significant inconsistency between the evidence and the
reference CWE. Only mark true when the evidence strongly
suggests the reference CWE is likely wrong, misleading, or
materially less supported than the predicted CWE.

Decision policy:\\
\hspace*{1em}If Situation~1 is true, decision = count\_as\_correct.\\
\hspace*{1em}Else if Situation~2 is true, decision = count\_as\_correct.\\
\hspace*{1em}Else if Situation~3 is true, decision = discard.\\
\hspace*{1em}Else decision = keep\_as\_incorrect.

Guidance: be fair about accepting alternative-but-defensible
views. Many CVEs can reasonably be labeled from different
perspectives (root cause, direct bug type, memory effect, injection
vector, access-control consequence). Do not count cases where the
prediction is only loosely related or merely adjacent in the
taxonomy.

Return strict JSON:\\
\{\\
\hspace*{1em}"situation\_1\_overlap\_ambiguity": true$|$false,\\
\hspace*{1em}"situation\_2\_alternative\_but\_defensible": true$|$false,\\
\hspace*{1em}"situation\_3\_reference\_inconsistency": true$|$false,\\
\hspace*{1em}"decision": "count\_as\_correct" $|$ "discard" $|$
"keep\_as\_incorrect",\\
\hspace*{1em}"reason": "short explanation"\\
\}

Evidence:\\
<<<EVIDENCE\\
<CVE description, advisory text, code/patch snippets, CodeQL
findings, Phase 2 reasoning>\\
EVIDENCE>>>
\end{tcolorbox}
\caption{Prompt used by the Arbitrator to adjudicate a
single (CVE, prediction, reference) tuple. The Arbitrator is
implemented with Claude Sonnet~4.6, a different LLM family from
the GPT-5.4-mini used in Phase~2. The prompt has been lightly
reflowed for the figure; the production version is available in
the artifact.}
\label{fig:agent3-prompt}
\end{figure}

\vspace{-6pt}
\section{{\tech} Implementation}
\label{sec:design:implementation}
\vspace{-6pt}

{\tech} uses three LLMs, chosen for the workload of each stage. Phase~1 uses Claude~4.6~Opus for seclometry generation and multi-view semantic clustering, because this phase performs one-time catalog-wide reasoning where quality matters more than per-call cost. Phase~2 uses GPT-5.4-mini for the Seclometry Builder and CWE Matcher, which run once per CVE and must scale to wild-scan measurement. The Arbitrator uses Claude Sonnet~4.6, providing model-family separation from the Classifier while remaining practical at measurement scale.

All LLM calls use fixed prompts, fixed model versions, and temperature~0, so each call is deterministic given its input. Deterministic post-processing enforces only output-format constraints, such as repairing malformed JSON, and does not alter the semantic content of any decision. The evidence-refinement loop between Agents~1 and~2 is capped at three iterations, an empirically chosen budget beyond which the predicted CWE rarely changes.

The Seclometry Builder uses standard web-fetching libraries to retrieve advisories and source artifacts, and invokes the CodeQL CLI with standard code-scanning query packs for the supported languages: C/C++, Python, JavaScript, Java, C\#, Go, and Ruby. {\tech} draws on three external data sources: the CWE catalog is MITRE's \emph{cwec\_latest.xml}, filtered to entries with \textsc{Allowed} mapping status (842 CWEs at the time of our study); CVE records come from the \emph{cvelistV5} Git repository at each CVE's pre-CWE-assignment commit, so the descriptions and references available to {\tech} precede the reference CWE assignment; and reference CWE labels and change histories are retrieved from NVD. To reduce data leakage, advisory and reference pages are fetched from Wayback Machine snapshots \textit{dated before the relevant CWE assignment}, or \textit{before the model's training cutoff when appropriate}, whenever such snapshots are available; live URLs are used only when no suitable snapshot exists.

\vspace{-0pt}
\section{Evaluation}
\label{sec:eval}
\vspace{-0pt}
We evaluate the two tools in {\tech}
separately, as each has its own
notion of correctness and no single dataset has both a known-correct
label and a known-incorrect existing label for the same CVE. The
Classifier is evaluated for prediction accuracy on CVEs whose
correct CWE is known (\textbf{RQ1}), and the Arbitrator for adjudication
reliability against CVEs whose existing-label correctness is known
(\textbf{RQ2}). %Validating each in isolation is what licenses composing them into the large-scale audit.

\vspace{-6pt}
\subsection{Classifier Prediction Accuracy (RQ1)}
\label{sec:eval:controlled}
\vspace{-6pt}
\textbf{Ground-truth data.} We built a high-confidence benchmark by
\emph{filtering} rather than re-labeling: two authors independently
inspected candidate CVEs---reading the description, advisory, and patch---and retained only those for which both
agreed that the NVD-assigned CWE clearly captures the root-cause weakness,
discarding any inaccurate, ambiguous, or contested label. The
ground-truth labels are thus NVD's own, which avoids author-side
labeling bias.

The resulting benchmark contains 100 CVEs from open-source projects
across 2017--2026 (at least three per year), spanning 38 distinct CWEs
with no CWE %exceeding 
covering more than 9\% of cases. 19 CVEs were disclosed after the
August~31, 2025 training cutoff of the LLM used by Agents~1 and~2, which
we keep as a leakage-mitigation subset.
We report \emph{top-1 exact-match accuracy}
(prediction equals ground-truth CWE) and \emph{ambiguity-aware accuracy}
(i.e., S1 + S2 + S3 in \S\ref{sec:design:agent3}). 
%; the two separate the strict matching rate from the rate of expert-defensible labels. 
% Here the ambiguity-aware judgment is made by manual adjudication,
% not the Arbitrator, since using the tool RQ2 evaluates to define
% ground-truth here would conflate the two questions.
%
To avoid conflating classifier evaluation with arbitrator evaluation, 
we use manual adjudication, not the Arbitrator, to compute ambiguity-aware accuracy here.

% \textbf{Baselines.} To separate {\tech}'s design from raw LLM
% capability, we compare against single-prompt baselines that start from
% the same per-CVE sources and the same 842-CWE candidate space but cannot
% invoke tools---they receive the CVE description and its referenced pages
% and predict in one call, whereas {\tech} actively gathers evidence
% (web search, webpage reader, source downloader, CodeQL) and runs its
% two-stage pipeline. We instantiate this baseline with two models:
% GPT-5.4-mini, the model {\tech} itself uses for Agents~1 and~2, and
% Claude~4.6~Opus, a substantially larger one. We additionally compare
% against TreeVul~\cite{pan2023fine}, a supervised classifier that predicts
% a CWE from a commit's code diff over a tree-structured model of the CWE
% hierarchy; we retrained it on its released dataset and evaluated it on
% our benchmark.

\textbf{Baselines.}
To separate {\tech}'s design from raw LLM
capability, we compare {\tech} against three baselines. The first two are single-prompt LLM baselines over the same 842-CWE candidate space: GPT-5.4-mini, the model used by {\tech}'s Agents~1 and~2, and Claude~4.6~Opus, a larger model. Both receive the same initial CVE sources but must predict in one call, without evidence-gathering tools or the two-stage matching pipeline. The third baseline is TreeVul~\cite{pan2023fine}, a SOTA (supervised diff-based) CWE classifier over the CWE hierarchy; we %re
train it on its released dataset and evaluate it on our benchmark.

% \textbf{Results.} {\tech} attains the highest accuracy on both metrics 
% (Table~\ref{tab:rq1}). %, and the two
% %single-prompt baselines isolate the source of {\tech}'s advantage.
% Against GPT-5.4-mini---the same model, evidence and model held
% constant---the large exact-match gap reflects the two-stage architecture
% and evidence-refinement loop. Against Claude~4.6~Opus, {\tech} leads by
% 11 points \emph{while using the smaller model}, so the gain is
% architectural rather than a function of model strength: one call to a
% stronger model still trails multi-view filtering, per-candidate scoring,
% and iterative refinement.
% TreeVul, despite being trained specifically for CWE prediction, trails
% even the same-model GPT-5.4-mini baseline: a supervised diff-only
% classifier over a restricted label space does not match an
% evidence-gathering pipeline over the full CWE catalog.

\begin{table}[tp]
\centering
\small
\caption{Accuracy and efficiency on the 100-CVE controlled benchmark.
\emph{Amb.}: ambiguity-aware
accuracy; \emph{Lat.}: per-CVE
latency in seconds; \emph{Cost}: per-CVE LLM cost in US dollars;
\emph{In/Out}: per-CVE input/output tokens in thousands}
\label{tab:rq1}
\setlength{\tabcolsep}{3pt}
\renewcommand{\arraystretch}{0.95}
\footnotesize
\scalebox{0.9}{
\begin{tabular}{lccrrr}
\hline
\textbf{System} & \textbf{Exact} & \textbf{Amb.} &
\textbf{Lat.\,(s)} & \textbf{Cost (\$)} & \textbf{In/Out (K)} \\
\hline
\tech                          & \textbf{85\%} & \textbf{92\%} & 61.7 & 0.070 & 76.6/2.8 \\
Claude 4.6 Opus (single)       & 74\% & 83\% & 51.7 & 0.172 & 14.9/3.9 \\
GPT-5.4-mini (single)          & 46\% & 59\% & 5.4  & 0.030 & 14.9/4.2 \\
TreeVul (diff, supervised)     & 52\% & 65\% & 3.7  & --- & --- \\
\hline
\end{tabular}}
\end{table}

\textbf{Results.} {\tech} achieves the highest exact-match (85\%) and ambiguity-aware (92\%) accuracy (Table~\ref{tab:rq1}). The efficiency numbers show where the improvement comes from. Compared with the GPT-5.4-mini single-prompt baseline---same base model, same evidence---{\tech} uses substantially more input context, reflecting Agent~1's evidence gathering and the per-candidate comparisons in Agent~2. The accuracy gain is therefore not attributable to model scale alone, but to converting CWE prediction into evidence construction followed by constrained semantic matching. Claude~4.6~Opus shows that a stronger single model can improve over GPT-5.4-mini yet still falls short of {\tech} while costing more per CVE. TreeVul's low cost and latency reflect the merits of supervised diff-based classification, but its lower accuracy exposes the limitation of restricted-label, diff-only prediction for full-catalog CWE auditing. Overall, {\tech}'s per-CVE cost is dominated by Agent~1's evidence gathering---roughly five times the input tokens of the single-prompt baselines---but remains cheaper per CVE than the Claude~4.6~Opus baseline while delivering substantially higher accuracy.

\vspace{-6pt}
\subsection{Arbitrator Adjudication Reliability (RQ2)}
\label{sec:rq2}
\vspace{-8pt}
To evaluate %\tech's 
the Arbitrator, we measure its precision and recall
across its three decision categories %(\textit{count\_as\_correct},
%\textit{discard}, and \textit{keep\_as\_incorrect}) 
on a unified test set of
100~CVEs constructed from two complementary sources.

\textbf{Test set.} The test set combines two
ground-truth halves independent of \tech. The first is 50~CVEs whose
assigned CWE was later revised in NVD's change history and which we
manually verified as semantically inconsistent with the evidence (true
NVD mislabels). The second is 50~CVEs sampled from the wild-scan
disagreements for which we manually confirmed NVD's label is correct
(cases the Arbitrator should not flag). Two authors independently labeled
each as \textit{count\_as\_correct}, % (S1+S2), 
\textit{discard} (NVD
inconsistent), or \textit{keep\_as\_incorrect} (\tech\ wrong), with
Cohen's $\kappa=0.84$; disagreements were resolved by discussion.

% \textbf{Results.} The Arbitrator achieves 82\% precision and 84\% recall
% in every category for an overall 90\% accuracy, with per-category F1
% tightly clustered (89--91\%)---its reliability is balanced, not skewed
% toward one class (full per-category breakdown in
% Table~\ref{tab:agent3-metrics}). A
% single LLM call over the already-gathered evidence makes it cheap enough
% to apply at the scale of the wild measurement.

\textbf{Results.}
The Arbitrator achieves 90\% overall accuracy, with per-category F1 scores tightly clustered between 89\% and 91\% (Table~\ref{tab:agent3-metrics}). Precision and recall are both at least 82\% in every category, indicating balanced reliability rather than performance concentrated on one outcome. Because adjudication is a single LLM call over evidence already gathered by the Classifier, it is inexpensive enough to apply at wild-measurement scale.

\begin{table}[tp]
\centering
\caption{Per-category and aggregate precision, recall, and F1 for
\tech's Arbitrator on the 100-CVE dataset}
\label{tab:agent3-metrics}
\setlength{\tabcolsep}{10pt}
\renewcommand{\arraystretch}{0.95}
\footnotesize
\scalebox{1.0}{
\begin{tabular}{lccc}
\hline
\textbf{Category} & \textbf{Precision} & \textbf{Recall} & \textbf{F1} \\
\hline
\textit{count\_as\_correct}   & 82.61\% & 97.44\% & 89.41\% \\
\textit{discard}              & 97.67\% & 84.00\% & 90.32\% \\
\textit{keep\_as\_incorrect}  & 90.91\% & 90.91\% & 90.91\% \\
\hline
Macro avg                     & 90.40\% & 90.78\% & 90.21\% \\
Micro avg                     & 90.00\% & 90.00\% & 90.00\% \\
Weighted avg                  & 91.05\% & 90.00\% & 90.03\% \\
\hline
\end{tabular}}
\end{table}

For the Arbitrator, the most important observation is the absence of a dominant failure mode. The high recall of \emph{count\_as\_correct} means the Arbitrator rarely rejects defensible labels, while the high precision of \emph{discard} means it is conservative when marking a reference label unreliable. This balance matters for our measurement: an overly aggressive Arbitrator would inflate the mislabel rate, whereas an overly permissive one would hide evidence-inconsistent labels. The clustered F1 scores indicate that neither effect dominates.

\vspace{-0pt}
\section{Measuring CWE Mislabeling in the NVD}
\label{sec:wild}
\vspace{-0pt}

We applied {\tech} to 15{,}556~CVEs disclosed between 2017 and 2026 with GitHub fix (patch) commits, and used the Arbitrator to adjudicate every disagreement between the Classifier's prediction and current NVD
label.

\vspace{-6pt}
\subsection{The Landscape of CWE Mislabels}
\label{sec:wild:landscape}
\vspace{-6pt}

\subsubsection{Overall Prevalence}
\label{sec:wild:prevalence}
Table~\ref{tab:landscape} summarizes the outcome distribution of the
15{,}556 adjudicated CVEs. {\tech} and NVD agree exactly in 49.70\% of
cases; a further 25.01\% are S2 overlap ambiguities and 6.36\% are S3
defensible alternatives (\S\ref{sec:design:agent3}), so NVD's label is
\emph{ambiguity-aware correct} in 81.07\%. The remaining disagreements
split into 3.63\% S4 verdicts, where NVD's label is inconsistent with the
evidence, and 15.30\% true classifier errors, where {\tech} is wrong and
NVD correct.

\begin{table}[t]
\centering
%\caption{Overall outcome distribution for the 15{,}556-CVE wild scan.}
\caption{Overall distribution of measurement outcomes}
\label{tab:landscape}
\setlength{\tabcolsep}{10pt}
\renewcommand{\arraystretch}{0.95}
\footnotesize
\begin{tabular}{lrr}
\hline
\textbf{Outcome} & \textbf{Count} & \textbf{Share} \\
\hline
%Exact match  
S1: exact match & 7{,}732 & 49.70\% \\
S2: overlap ambiguity                & 3{,}890 & 25.01\% \\
S3: defensibly alternative           &    990 &  6.36\% \\
S4: NVD inconsistent with evidence   &    564 &  3.63\% \\
Classifier Error (\textbf{Err}): NVD label correct    & 2{,}380 & 15.30\% \\
\hline
\textbf{Ambiguity-aware} (\textbf{AmbAw}; S1$+$S2$+$S3) & 12{,}612 & 81.07\% \\
\hline
\end{tabular}
\vspace{-0pt}
\end{table}

% Two facts frame the analysis that follows. First, the dominant source
% of label--prediction mismatch is not labeler error but taxonomy
% overlap: the bulk of the 50.30\% mismatch is defensibly-equivalent
% (S1/S2) labeling rather than inconsistency. Second, the evidence-%
% inconsistent tail, though only 3.63\% of the corpus, is a substantial
% absolute count (564 CVEs) and is where genuine mislabels concentrate.

% \vspace{-4pt}
% \begin{tcolorbox}[left=1pt,right=1pt,top=0pt,bottom=2pt,lifted shadow={1mm}{-2mm}{3mm}{0.1mm}{black!50!white},arc=0pt,auto outer arc,boxrule=.5pt,leftrule=2pt]
% \textbf{Finding 1:} NVD's CWE label mismatches the strict
% evidence-supported assignment on half the corpus, but most mismatches
% are taxonomy overlap, not error: only 3.63\% of labels are
% evidence-inconsistent. NVD labels are far more often imprecise than wrong.
% \vspace{-4pt}
% \end{tcolorbox}

\vspace{-6pt}
\subsubsection{Longitudinal Trend}
\label{sec:wild:temporal}

Table~\ref{tab:per-year} reports outcome rates by CVE publication year
across the full ten-year span (2017--2026). NVD's strict exact-match
rate is essentially \emph{stable} across the period, ranging from
43.72\% (2019) to 53.76\% (2018) with no monotonic trend. By this
strictest measure, NVD's CWE labels are no worse on CVEs published in
2026 than on those from 2017.

The picture changes when the disagreement cases are examined more
carefully. The share of CVEs carrying CWE labels that are inconsistent
with their own evidence rises substantially: from 1--3\% in 2017--2018
to 3--6\% in 2021--2026. In absolute
terms, the 2017--2018 cohort contributes 27 inconsistent labels, while
the 2024--2026 cohort contributes 228, accounting for over 40\% of
all evidence-inconsistent labels in the corpus.
The share of CVEs whose labels are neither most-specific nor
defensibly equivalent also rises monotonically from 9.7--12.9\% in
2017--2020 to 14.5--18.7\% in 2022--2026, contributing to a 6--9
percentage point decline in ambiguity-aware accuracy across the
period (86--88\% in 2017--2018 versus 78--81\% in 2023--2026).
Together these indicate a decline in CWE labeling quality for
open-source CVEs as the annual volume of disclosed CVEs has continued to
grow~\cite{nistGrowth2026}.

\begin{table}[t]
\centering
\caption{Outcome distribution by CVE publication year 
%AmbAw is the ambiguity-aware rate (S1 $+$ S2 $+$ S3).
}
\label{tab:per-year}
\setlength{\tabcolsep}{6pt}
\renewcommand{\arraystretch}{0.95}
\footnotesize
\rowcolors{2}{gray!15}{white}
\scalebox{0.9}{
\begin{tabular}{lrrrrr}
\hline
\textbf{Year} & \textbf{N} & \textbf{S1\%} & \textbf{AmbAw\%} &
\textbf{S4\%} & \textbf{Err\%} \\
\hline
2017 &    911 & 50.93 & 86.06 & 0.99 & 12.95 \\
2018 &    731 & 53.76 & 87.55 & 2.46 &  9.99 \\
2019 &    693 & 43.72 & 84.27 & 6.06 &  9.67 \\
2020 &    985 & 51.07 & 85.28 & 2.94 & 11.78 \\
2021 & 1{,}626 & 52.40 & 80.69 & 4.80 & 14.51 \\
2022 & 2{,}071 & 52.00 & 80.15 & 3.67 & 16.18 \\
2023 & 2{,}021 & 51.36 & 78.38 & 4.16 & 17.47 \\
2024 & 2{,}074 & 47.54 & 81.05 & 2.85 & 16.10 \\
2025 & 2{,}179 & 46.08 & 80.54 & 4.54 & 14.91 \\
2026 & 2{,}265 & 49.09 & 78.23 & 3.09 & 18.68 \\
\hline
Total & 15{,}556 & 49.70 & 81.07 & 3.63 & 15.30 \\
\hline
\end{tabular}}
\vspace{-0pt}
\end{table}

\vspace{-6pt}
\begin{tcolorbox}[left=1pt,right=1pt,top=0pt,bottom=2pt,lifted shadow={1mm}{-2mm}{3mm}{0.1mm}{black!50!white},arc=0pt,auto outer arc,boxrule=.5pt,leftrule=2pt]
\textbf{Finding 1:} NVD's CWE labels are far more often imprecise than
wrong: half the corpus mismatches the strict evidence-supported
assignment, but most mismatches are taxonomy overlap and only 3.63\% are
evidence-inconsistent. Exact-match accuracy is stable across 2017--2026,
yet the evidence-inconsistent share has grown, indicating a decline in
labeling quality as CVE volume rises.
\vspace{-12pt}
\end{tcolorbox}

\vspace{-6pt}
\subsubsection{Distribution}
\label{sec:cwe-dist}
We now decompose label quality along four axes---the weakness type
itself, the project, the programming language, and the assigning
CNA---to locate where mislabeling concentrates.

\textbf{Over CWE.} 
We group the
15{,}556 CVEs by their NVD-assigned CWE's parent category in the
CWE-699 software development view, and separately by their specific
CWEs. %The per-family distribution is reported in
%Table~\ref{tab:cwe-family}, and the top specific CWEs by volume are
%summarized in Table~\ref{tab:cwe-top}.

Across CWE-699 categories (Table~\ref{tab:cwe-family}), exact-match
varies more than four-fold while evidence-inconsistency stays in a narrow
band. At the family level NVD's labels thus differ enormously in
\emph{precision} but only modestly in \emph{correctness}: a family whose
labels are rarely most-specific (e.g., authorization) is not thereby
evidence-inconsistent. Low family-level exact-match reflects abstract or
defensibly-alternative labeling, not outright error.

% \vspace{-4pt}
% \begin{tcolorbox}[left=1pt,right=1pt,top=0pt, bottom=2pt, lifted shadow={1mm}{-2mm}{3mm}{0.1mm}{black!50!white},arc=0pt,auto outer arc, boxrule=.5pt,leftrule=2pt]
% \textbf{Finding 3:} At the CWE-family level, NVD's labels vary
% four-fold in exact-match precision (over 80\% for injection down to
% under 20\% for authorization) but vary only modestly in
% evidence-inconsistency (roughly 2--6\% S3 across all large families).
% Low family-level exact-match reflects abstract or defensibly-alternative
% labeling, not outright error.
% \vspace{-4pt}
% \end{tcolorbox}

The family-level band conceals sharp variation among specific
CWEs (Table~\ref{tab:cwe-top}). Heap-based buffer overflow (CWE-122) is
an outlier at 11.94\% evidence-inconsistency---over triple the
corpus rate---while the highest-volume CWEs are among the most
consistent. Because CWE-122 is within the larger memory-buffer
family, this outlier is averaged away at the family level: genuine NVD
mislabels localize to specific weakness types, not whole families.

Two groups have near-zero exact-match by construction: discouraged
abstract CWEs (e.g., CWE-20, CWE-200) that MITRE prohibits as primary
labels, and CWEs with multiple valid nesting levels (e.g., use-after-free
CWE-416), which the evidence supports defensibly but rarely at the most-%
specific level (Table~\ref{tab:cwe-top}). In both, the labels are
predominantly not \emph{wrong}---S4 at or below the corpus average---but
imprecise, leaving users to resolve the actual weakness.

% \vspace{-4pt}
% \begin{tcolorbox}[left=1pt,right=1pt,top=0pt, bottom=2pt, lifted shadow={1mm}{-2mm}{3mm}{0.1mm}{black!50!white},arc=0pt,auto outer arc, boxrule=.5pt,leftrule=2pt]
% \textbf{Finding 4:} NVD's genuine mislabels localize to specific CWEs
% rather than whole families. Heap-based buffer overflow (CWE-122) is the
% clear outlier at 11.94\% evidence-inconsistency, over triple the
% corpus-wide 3.63\%, while high-volume CWEs such as cross-site scripting
% (1.52\%) are highly evidence-consistent.
% \vspace{-4pt}
% \end{tcolorbox}

\begin{table}[t]
\centering
\caption{Outcome distribution by CWE-699 category (categories with
$\geq$100 CVEs)}
\label{tab:cwe-family}
\small
\setlength{\tabcolsep}{4pt}
\renewcommand{\arraystretch}{0.95}
\rowcolors{2}{gray!15}{white}
\scalebox{0.7}{
\begin{tabular}{lrrrrr}
\hline
\textbf{Category} & \textbf{N} & \textbf{S1\%} &
\textbf{AmbAw\%} & \textbf{S4\%} & \textbf{Err\%} \\
\hline
CWE-137  Data Neutralization        & 3689 & 81.92 & 91.41 & 2.11 &  6.29 \\
CWE-1218 Memory Buffer Errors       & 1160 & 58.36 & 90.52 & 3.63 &  5.78 \\
CWE-1219 File Handling              &  878 & 69.59 & 89.29 & 2.51 &  8.09 \\
CWE-399  Resource Management        &  593 & 53.79 & 80.78 & 4.89 & 14.33 \\
CWE-465  Pointer Issues             &  459 & 78.00 & 88.45 & 2.61 &  8.93 \\
CWE-840  Business Logic Errors      &  432 & 34.95 & 65.28 & 4.17 & 30.56 \\
CWE-417  Communication Channel      &  419 & 80.19 & 89.26 & 2.63 &  8.11 \\
CWE-189  Numeric Errors             &  401 & 67.83 & 84.54 & 3.24 & 11.97 \\
CWE-19   Data Processing            &  312 & 68.27 & 78.21 & 3.85 & 17.95 \\
CWE-1211 Authentication Errors      &  266 & 26.32 & 67.67 & 4.51 & 27.07 \\
CWE-389  Error Conditions           &  230 & 55.22 & 76.09 & 4.35 & 18.70 \\
CWE-199  Information Management     &  213 & 37.09 & 82.16 & 3.76 & 13.15 \\
CWE-1212 Authorization Errors       &  201 & 18.41 & 45.77 & 2.99 & 51.24 \\
CWE-1214 Data Integrity             &  197 & 31.98 & 75.13 & 4.57 & 20.30 \\
CWE-438  Behavioral Problems        &  169 & 55.62 & 75.74 & 4.14 & 20.12 \\
CWE-429  Handler Errors             &  157 & 69.43 & 84.71 & 5.73 &  9.55 \\
CWE-310  Cryptographic Issues       &  157 & 32.48 & 80.25 & 2.55 & 17.20 \\
CWE-1226 Complexity Issues          &  145 & 73.79 & 86.90 & 3.45 &  9.66 \\
CWE-255  Credentials Management     &  106 & 50.00 & 78.30 & 3.77 & 16.98 \\
\hline
\end{tabular}}
\vspace{-0pt}
\end{table}

\begin{table}[t]
\centering
\caption{Outcome distribution by specific CWE (top 20)}
\label{tab:cwe-top}
\small
\setlength{\tabcolsep}{4pt}
\renewcommand{\arraystretch}{0.95}
\rowcolors{2}{gray!15}{white}
\scalebox{0.7}{
\begin{tabular}{lrrrrr}
\hline
\textbf{CWE} & \textbf{N} & \textbf{S1\%} &
\textbf{AmbAw\%} & \textbf{S4\%} & \textbf{Err\%} \\
\hline
CWE-79  XSS                        & 2506 & 89.07 & 93.26 & 1.52 &  5.07 \\
CWE-22  Path Traversal             &  741 & 72.60 & 91.23 & 2.29 &  6.34 \\
CWE-125 Out-of-bounds Read         &  635 & 77.64 & 92.44 & 2.05 &  5.35 \\
CWE-20  Improper Input Validation  &  461 &  0.00 & 65.29 & 2.60 & 30.80 \\
CWE-200 Info Exposure              &  448 &  0.00 & 75.45 & 4.46 & 19.64 \\
CWE-89  SQL Injection              &  441 & 83.90 & 90.02 & 2.95 &  7.03 \\
CWE-476 NULL Pointer Deref         &  409 & 84.60 & 92.18 & 2.44 &  5.38 \\
CWE-918 SSRF                       &  361 & 85.87 & 90.58 & 2.77 &  6.65 \\
CWE-787 Out-of-bounds Write        &  344 & 47.09 & 88.08 & 6.10 &  5.81 \\
CWE-78  OS Command Injection       &  325 & 61.85 & 88.31 & 3.38 &  8.00 \\
CWE-400 Resource Exhaustion        &  314 &  0.00 & 71.02 & 5.10 & 23.57 \\
CWE-416 Use After Free             &  298 & 21.81 & 93.29 & 3.36 &  3.36 \\
CWE-352 CSRF                       &  277 & 85.20 & 88.45 & 3.61 &  7.94 \\
CWE-94  Code Injection             &  276 & 50.36 & 82.97 & 4.35 & 11.96 \\
CWE-863 Incorrect Authorization    &  250 & 26.40 & 58.00 & 1.20 & 39.60 \\
CWE-770 Resource Allocation        &  241 & 46.47 & 82.57 & 4.56 & 12.86 \\
CWE-862 Missing Authorization      &  232 & 46.98 & 62.50 & 2.16 & 35.34 \\
CWE-190 Integer Overflow           &  230 & 68.26 & 83.91 & 1.74 & 13.91 \\
CWE-122 Heap-based Buffer Overflow &  201 & 44.28 & 79.60 &11.94 &  7.96 \\
CWE-284 Improper Access Control    &  200 &  0.00 & 55.50 & 4.50 & 39.00 \\
\hline
\end{tabular}}
\vspace{-0pt}
\end{table}

\textbf{Over Project.} 
We identify the project for 15{,}178 of 15{,}556 CVEs (97.6\%) from the
fix-commit URL (Table~\ref{tab:projects}).
Project-level exact-match varies widely, but
the split is not random: injection-heavy web applications are labeled
precisely while memory-safety-heavy systems software is labeled
imprecisely. Projects inherit the precision of whatever CWE families
dominate their vulnerabilities---low project-level exact-match is largely
a restatement of the memory-safety precision gap, not an independent
project effect.

% \vspace{-4pt}
% \begin{tcolorbox}[left=1pt,right=1pt,top=0pt, bottom=2pt, lifted shadow={1mm}{-2mm}{3mm}{0.1mm}{black!50!white},arc=0pt,auto outer arc, boxrule=.5pt,leftrule=2pt]
% \textbf{Finding 5:} Project-level precision differences are a composition effect---projects inherit the precision of the weakness types that dominate them, not an independent project signal.
% \vspace{-4pt}
% \end{tcolorbox}

Evidence-inconsistency, by contrast, is uniform and near the corpus
average across almost all projects, with one exception: Radare2, whose
15.49\% S4 rate is over four times the corpus average and the highest in
the corpus---a genuine concentration of NVD labels that contradict the
CVEs' own evidence, not an artifact of low precision.

% \vspace{-4pt}
% \begin{tcolorbox}[left=1pt,right=1pt,top=0pt, bottom=2pt, lifted shadow={1mm}{-2mm}{3mm}{0.1mm}{black!50!white},arc=0pt,auto outer arc, boxrule=.5pt,leftrule=2pt]
% \textbf{Finding 6:} Evidence-inconsistency is low and uniform across
% nearly all projects (below 6\%, near the corpus average), with one
% exception: Radare2's 15.49\% S3 rate is over four times the corpus
% average, indicating a project-specific concentration of NVD labels
% that contradict the CVEs' own evidence.
% \vspace{-4pt}
% \end{tcolorbox}

\begin{table}[t]
\centering
\caption{Outcome distribution by project (top 20)}
\label{tab:projects}
\small
\setlength{\tabcolsep}{6pt}
\renewcommand{\arraystretch}{0.95}
\rowcolors{2}{gray!15}{white}
\scalebox{0.7}{
\begin{tabular}{lrrrrr}
\hline
\textbf{Project} & \textbf{N} & \textbf{S1\%} &
\textbf{AmbAw\%} & \textbf{S4\%} & \textbf{Err\%} \\
\hline
tensorflow/tensorflow      & 390 & 53.59 & 81.03 & 1.54 & 17.18 \\
torvalds/linux             & 363 & 30.58 & 82.37 & 4.41 & 13.22 \\
vim/vim                    & 186 & 52.69 & 87.10 & 4.30 &  8.60 \\
xwiki/xwiki-platform       & 181 & 43.65 & 77.35 & 3.31 & 16.57 \\
gpac/gpac                  & 150 & 46.67 & 88.00 & 5.33 &  6.67 \\
imagemagick/imagemagick    & 149 & 40.27 & 83.89 & 2.68 & 13.42 \\
tcpdump-group/tcpdump      & 111 & 86.49 & 97.30 & 0.90 &  1.80 \\
discourse/discourse        & 108 & 28.70 & 72.22 & 0.93 & 26.85 \\
openemr/openemr            & 102 & 58.82 & 78.43 & 0.98 & 20.59 \\
openclaw/openclaw          & 102 & 44.12 & 79.41 & 0.98 & 19.61 \\
ffmpeg/ffmpeg              &  98 & 39.80 & 81.63 & 6.12 & 12.24 \\
freerdp/freerdp            &  94 & 54.26 & 91.49 & 2.13 &  5.32 \\
thorsten/phpmyfaq          &  89 & 83.15 & 91.01 & 0.00 &  8.99 \\
pimcore/pimcore            &  87 & 85.06 & 94.25 & 2.30 &  3.45 \\
misp/misp                  &  86 & 69.77 & 84.88 & 1.16 & 13.95 \\
microweber/microweber      &  80 & 60.00 & 80.00 & 3.75 & 16.25 \\
radareorg/radare2          &  71 & 38.03 & 78.87 &15.49 &  5.63 \\
wwbn/avideo                &  67 & 73.13 & 95.52 & 0.00 &  4.48 \\
librenms/librenms          &  61 & 91.80 & 93.44 & 0.00 &  6.56 \\
\hline
\end{tabular}}
\vspace{-2pt}
\end{table}

\textbf{Over Programming Languages.} Label quality also differs across languages (Table~\ref{tab:languages}), but the per-language CWE breakdown shows this is a composition effect: a given CWE is labeled with similar quality regardless of language, so languages that appear well- or poorly-labeled in aggregate are simply those dominated by precisely- or imprecisely-labeled weakness types (injection versus memory-safety). Both S1 and S4 echo the weakness-type pattern, revealing no language-intrinsic effect.

The extremes illustrate the two sources of aggregate variation. PHP has the highest S1 rate (66.02\%) and one of the highest ambiguity-aware rates (85.17\%), consistent with a corpus dominated by well-established web-vulnerability categories where NVD labels, advisories, and CWE definitions use standardized language. Rust has the lowest S1 rate (31.97\%), lowest ambiguity-aware rate (70.49\%), and highest S4 rate (6.15\%): the Rust CVEs that remain visible in NVD often involve unsafe code, FFI boundaries, parser/runtime behavior, or resource exhaustion---cases that sit near boundaries between broad CWE classes and lack stable labeling conventions. C/C++ shows a third pattern: despite the largest sample size and a low exact-match rate (42.51\%), its ambiguity-aware rate remains high (82.07\%), indicating that many memory-safety disagreements are granularity or boundary disputes---e.g., an out-of-bounds read vs.\ write, or buffer overflow vs.\ broader memory corruption---rather than outright inconsistencies. This supports the paper's central distinction between imprecision and mislabeling.

\begin{table}[t]
\vspace{-0pt}
\centering
\caption{Outcome distribution by language ($\geq$100 CVEs)}
\label{tab:languages}
\small
\setlength{\tabcolsep}{10pt}
\renewcommand{\arraystretch}{0.95}
\rowcolors{2}{gray!15}{white}
\scalebox{0.72}{
\begin{tabular}{lrrrrr}
\hline
\textbf{Language} & \textbf{N} & \textbf{S1\%} &
\textbf{AmbAw\%} & \textbf{S4\%} & \textbf{Err\%} \\
\hline
C/C++      & 4484 & 42.51 & 82.07 & 4.42 & 13.16 \\
PHP        & 3305 & 66.02 & 85.17 & 2.57 & 12.07 \\
JavaScript & 1715 & 51.14 & 81.40 & 2.74 & 15.57 \\
Python     & 1557 & 49.71 & 82.79 & 3.21 & 13.55 \\
Go         &  987 & 39.72 & 76.09 & 3.65 & 19.76 \\
Java       &  755 & 47.55 & 81.99 & 2.91 & 14.04 \\
TypeScript &  726 & 48.76 & 79.06 & 3.58 & 16.67 \\
HTML       &  454 & 54.19 & 83.48 & 2.86 & 13.22 \\
Ruby       &  375 & 47.47 & 82.13 & 2.13 & 15.47 \\
Rust       &  244 & 31.97 & 70.49 & 6.15 & 22.95 \\
C\#        &  132 & 51.52 & 79.55 & 4.55 & 15.15 \\
Perl       &  104 & 50.96 & 83.65 & 0.00 & 16.35 \\
\hline
\end{tabular}}
\vspace{-2pt}
\end{table}

% \vspace{-4pt}
% \begin{tcolorbox}[left=1pt,right=1pt,top=0pt, bottom=2pt, lifted shadow={1mm}{-2mm}{3mm}{0.1mm}{black!50!white},arc=0pt,auto outer arc, boxrule=.5pt,leftrule=2pt]
% \textbf{Finding 7:} Apparent language-level differences in NVD label
% quality are a composition effect, not a language effect. Aggregate
% exact-match (66\% PHP to 32\% Rust) and S3 (2--6\%) both track the
% weakness types that dominate each language's corpus: injection-heavy
% languages are precise and evidence-consistent, memory-safety-heavy
% languages are imprecise and modestly more inconsistent.
% \vspace{-4pt}
% \end{tcolorbox}

\textbf{Over CNAs.}
Conditioning on the assigning CNA produces the widest quality spread in
the entire measurement: a roughly 70-point exact-match range
(Table~\ref{tab:cna}), far exceeding the variation across years, CWE
families, projects, or languages. The strongest performers (e.g.,
Wordfence, Snyk) combine high precision with near-zero
evidence-inconsistency, while the two highest-volume CNAs sit mid-range,
so high precision is not merely a small-specialist effect. The
trustworthiness of an NVD CWE label thus depends substantially on which
organization produced it.

The S4 spread across CNAs is narrower than the exact-match spread
(Table~\ref{tab:cna}), repeating the pattern seen throughout: organizations
differ most in how \emph{precisely} they label, less in how often they are
outright \emph{wrong}.

\begin{table}[t]
\centering
\caption{Outcome distribution by assigning CNA (top 15 by CVE count
among CNAs with $\geq$20 CVEs)}
\label{tab:cna}
\small
\setlength{\tabcolsep}{10pt}
\renewcommand{\arraystretch}{0.95}
\rowcolors{2}{gray!15}{white}
\scalebox{0.72}{
\begin{tabular}{lrrrrr}
\hline
\textbf{CNA} & \textbf{N} & \textbf{S1\%} &
\textbf{AmbAw\%} & \textbf{S4\%} & \textbf{Err\%} \\
\hline
github\_m    & 7{,}599 & 45.85 & 78.62 & 3.53 & 17.25 \\
mitre        & 3{,}574 & 48.57 & 84.75 & 4.14 & 11.02 \\
@huntrdev    & 1{,}526 & 63.11 & 78.77 & 4.46 & 16.58 \\
vuldb        &    684 & 66.37 & 86.84 & 3.51 &  9.36 \\
@huntr\_ai   &    403 & 49.13 & 75.68 & 2.23 & 21.09 \\
redhat       &    324 & 39.20 & 83.95 & 2.47 & 12.96 \\
snyk         &    313 & 58.15 & 88.50 & 1.92 &  8.95 \\
vulncheck    &    293 & 43.00 & 69.97 & 3.75 & 26.28 \\
mend         &    103 & 66.99 & 74.76 & 2.91 & 22.33 \\
facebook     &     88 & 48.86 & 79.55 & 3.41 & 14.77 \\
wordfence    &     79 & 83.54 & 91.14 & 0.00 &  8.86 \\
hackerone    &     54 & 57.41 & 92.59 & 0.00 &  7.41 \\
flexera      &     23 & 52.17 &100.00 & 0.00 &  0.00 \\
checkpoint   &     21 & 61.90 &100.00 & 0.00 &  0.00 \\
apache       &     20 & 55.00 & 95.00 & 0.00 &  5.00 \\
\hline
\end{tabular}}
\vspace{-5pt}
\end{table}

\vspace{-6pt}
\begin{tcolorbox}[left=1pt,right=1pt,top=0pt, bottom=2pt, lifted shadow={1mm}{-2mm}{3mm}{0.1mm}{black!50!white},arc=0pt,auto outer arc, boxrule=.5pt,leftrule=2pt]
\textbf{Finding 2:} Across all four axes, NVD labels vary far more in
\emph{precision} than in \emph{correctness}. Exact-match swings widely
by weakness type and by assigning CNA (a $\sim$70-point
spread), while evidence-inconsistency stays in a narrow band everywhere.
Project- and language-level differences are composition effects of the
underlying weakness types, not independent signals; genuine mislabels
localize to a few specific CWEs (led by heap buffer overflow) and, at
the project level, to Radare2.
\vspace{-2pt}
\end{tcolorbox}

% \vspace{-4pt}
% \begin{tcolorbox}[left=1pt,right=1pt,top=0pt, bottom=2pt, lifted shadow={1mm}{-2mm}{3mm}{0.1mm}{black!50!white},arc=0pt,auto outer arc, boxrule=.5pt,leftrule=2pt]
% \textbf{Finding 8:} The assigning CNA is the strongest predictor of
% label quality of any axis measured, with exact-match rates spanning
% roughly 70 points (13\% to 83\%) across CNAs. Since CNA identity is
% recorded in every CVE entry, it is the one quality signal a consumer
% can read directly off the metadata.
% \vspace{-4pt}
% \end{tcolorbox}
\vspace{-6pt}
\subsection{Severity and Impact}
\label{sec:wild:severity-impact}
\vspace{-6pt}
\subsubsection{CVE Severity}
\label{sec:wild:severity}

We first ask whether CVSS severity---the metadata that most directly
drives downstream prioritization---predicts label quality. Results are
in Table~\ref{tab:severity}.

Across the four CVSS severity bands, NVD's CWE label quality is essentially flat. Exact-match rates span less than six percentage points (Low at 54.46\% to High at 48.58\%) and evidence-inconsistency rates span less than one point (3.10\% to 3.78\%), with neither measure varying monotonically with severity. Critical-severity CVEs---the ones whose metadata most directly drives downstream prioritization and remediation---are labeled no more accurately than Medium-severity CVEs. CWE labeling is thus not being systematically refined for the vulnerabilities that are most severe or most likely to receive downstream attention.

A methodological implication is that severity cannot be used to obtain a cleaner CWE-labeled subset. If CWE-label quality tracked CVSS severity, researchers could reduce label noise by controlling for severity or by restricting analyses to High/Critical CVEs. Table~\ref{tab:severity} shows that this does not work: high-severity subsets retain nearly the same ambiguity-aware correctness and evidence-inconsistency rates as lower-severity subsets. Filtering by severity may change the risk profile of a dataset, but it does not materially improve the reliability of its CWE labels. Studies that depend on NVD-derived CWEs therefore need controls based on weakness semantics, evidence provenance, and assigning organization, rather than treating CVSS severity as a proxy for label trustworthiness.

\begin{table}[t]
\centering
\caption{Outcome distribution by CVSS severity range}
\label{tab:severity}
\small
\setlength{\tabcolsep}{10pt}
\renewcommand{\arraystretch}{0.95}
\scalebox{0.72}{
\begin{tabular}{lrrrrr}
\hline
\textbf{Severity} & \textbf{N} & \textbf{S1\%} &
\textbf{AmbAw\%} & \textbf{S4\%} & \textbf{Err\%} \\
\hline
Critical & 1{,}422 & 49.09 & 81.43 & 3.45 & 13.78 \\
High     & 4{,}343 & 48.58 & 79.67 & 3.78 & 16.05 \\
Medium   & 5{,}390 & 50.30 & 78.48 & 3.67 & 17.48 \\
Low      & 1{,}098 & 54.46 & 81.24 & 3.10 & 15.30 \\
\hline
\end{tabular}}
\vspace{-0pt}
\end{table}

% \vspace{-4pt}
% \begin{tcolorbox}[left=1pt,right=1pt,top=0pt, bottom=2pt, lifted shadow={1mm}{-2mm}{3mm}{0.1mm}{black!50!white},arc=0pt,auto outer arc, boxrule=.5pt,leftrule=2pt]
% \textbf{Finding 9:} CVSS severity does not predict NVD CWE label
% quality: exact-match rates span under six points and
% evidence-inconsistency rates under one point across all four
% severity bands, neither varying monotonically. Critical-severity
% CVEs are labeled no more accurately than Medium-severity ones.
% \vspace{-4pt}
% \end{tcolorbox}
\vspace{-6pt}
\subsubsection{Exploitability}
\label{sec:wild:exploitability}

Severity is a coarse proxy for a vulnerability's importance; whether a
CVE is actually \emph{exploited} is the sharpest one. We therefore ask
whether label quality differs for exploited CVEs, joining the corpus
against two independent exploitation signals: the CISA Known Exploited
Vulnerabilities (KEV) catalog~\cite{cisa-kev}, which records
vulnerabilities exploited in the wild, and Exploit-DB~\cite{exploitdb},
which records published proof-of-concept exploits.

Of the 15{,}556 CVEs, 241 have an Exploit-DB entry. Their
evidence-inconsistency rate is 3.73\% (9 of 241), statistically
indistinguishable from the 3.63\% rate among the remaining CVEs; their exact-match
(56.85\%) and ambiguity-aware (87.55\%) rates are likewise close to the
corpus as a whole. Only 39 corpus CVEs appear in KEV---too few to
support a standalone rate---but their S4 rate (2.56\%) is
consistent with the same picture.
As with severity, exploitation status does not predict label
trustworthiness.

\vspace{-4pt}
\begin{tcolorbox}[left=1pt,right=1pt,top=0pt, bottom=2pt, lifted shadow={1mm}{-2mm}{3mm}{0.1mm}{black!50!white},arc=0pt,auto outer arc, boxrule=.5pt,leftrule=2pt]
\textbf{Finding 3:} Neither CVSS severity nor real-world exploitation
predicts CWE label quality. A vulnerability's impact is not a proxy
for its label's trustworthiness.
\vspace{-2pt}
\end{tcolorbox}

% \vspace{-4pt}
% \begin{tcolorbox}[left=1pt,right=1pt,top=0pt, bottom=2pt, lifted shadow={1mm}{-2mm}{3mm}{0.1mm}{black!50!white},arc=0pt,auto outer arc, boxrule=.5pt,leftrule=2pt]
% \textbf{Finding 10:} Exploitation status does not predict NVD CWE
% label quality. Among the 241 corpus CVEs with a public Exploit-DB
% proof-of-concept, the evidence-inconsistency rate (3.73\%) matches the
% corpus-wide 3.62\% (Fisher's exact $p=0.86$); the 39 KEV-listed CVEs
% show the same. As with severity, a vulnerability's real-world impact
% does not make its CWE label more trustworthy.
% \vspace{-4pt}
% \end{tcolorbox}

\vspace{-6pt}
\subsection{Failure Patterns}
\label{sec:wild:failure-patterns}
\vspace{-6pt}
\subsubsection{Individual Label Failures}
\label{sec:qual-patterns}

We now examine the 564 S4 records, characterizing not their
number but \emph{how} the labels fail---what mismatch occurs between a
record's CWE and the weakness its evidence describes, and what property
of the record or taxonomy plausibly gives rise to it.

% Of the 564 S3 records, manual review confirmed 434 as genuine
% mislabels (the procedure is detailed below); all percentages in this
% section are computed over those 434. Across them we observe six
% recurring patterns, summarized in Table~\ref{tab:pattern-dist}. The
% patterns are not artifacts of any single project, language, or CWE
% family, and they are not equiprobable: the two largest---confusion
% between a weakness and its downstream consequence, and confusion among
% sibling sub-types within a family---together account for over
% two-fifths of confirmed mislabels, and both reflect how the CWE
% taxonomy is structured rather than project-specific mistakes.

% % ---------------------------------------------------------------------
% \textbf{Methodology.}

Two authors independently reviewed all 564 flagged records against their
primary sources in order of authority (fixing patch, advisory,
descriptions, proof-of-concept), confirming a record as a mislabel only
when the highest-authority evidence contradicted the assigned CWE and
returning defensible-but-imprecise readings to the defensible pool. This
confirmed 434 genuine mislabels and excluded 130 ($\kappa=0.87$); we
reported the confirmed set to NVD and received acknowledgment. To
characterize \emph{how} the labels fail, three authors then open-coded
the relationship between each assigned CWE and the evidence-supported
weakness, merging codes and applying the resulting codebook to all 434.
The codebook describes the records, not analyst intent; where we discuss
why a mismatch arises we point only to observable properties of the
record or taxonomy.

\begin{table}[t]
\centering
\caption{Distribution across confirmed mislabels}
\label{tab:pattern-dist}
\small
\setlength{\tabcolsep}{15pt}
\renewcommand{\arraystretch}{0.95}
\rowcolors{2}{gray!15}{white}
\scalebox{0.8}{
\begin{tabular}{lr}
\hline
\textbf{Pattern} & \textbf{Share} \\
\hline
Consequence vs.\ root cause            & 23.1\% \\
Sub-type mislabel (within family)      & 21.3\% \\
Injection sink / interpreter confusion & 19.5\% \\
Access-control \& identity conflation  & 15.3\% \\
Discouraged / wrong-branch label       & 10.3\% \\
Multi-CWE accumulation                 & 7.8\%  \\
\hline
(residual: file/path handling)         & 3.9\%  \\
\hline
\end{tabular}}
\vspace{-0pt}
\end{table}

% ---------------------------------------------------------------------
%\textbf{Patterns of Mislabeling.}

Open coding of the 434 confirmed mislabels revealed six recurring
patterns (Table~\ref{tab:pattern-dist}), each defined by how the
assigned CWE relates to the evidence-supported weakness. The top three
alone cover nearly two-thirds of the set, and every pattern turns on a
CWE-to-CWE relationship---cause vs.\ effect, sibling sub-types, sibling
sinks. We describe each
below with a representative example.
\textbf{Consequence vs.\ root cause (23.1\%).}
The most common pattern labels a vulnerability by a true \emph{outcome}
of the flaw rather than its \emph{root cause}, naming an effect
downstream on the causal chain. The motivating example of \S\ref{sec:bg:motiv} (CVE-2026-26014, Figure~\ref{fig:case-consequence}) is representative: nonce reuse (CWE-323) is the root cause but the record carries CWE-200 (information exposure), the consequence. Such broad consequence CWEs plausibly fit
the end state of almost any data-exposure flaw, so a label derived from
the described effect settles on them even when the description names the
mechanism.

\begin{figure}[t]
\begin{tcolorbox}[left=3pt,right=3pt,top=2pt,bottom=3pt,
  lifted shadow={1mm}{-2mm}{3mm}{0.1mm}{black!50!white},
  arc=0pt,auto outer arc,boxrule=.5pt,leftrule=2pt,
  title={\textbf{Case 2 (Sub-type mislabel): CVE-2022-1160}},
  fonttitle=\footnotesize, fontupper=\scriptsize]
\vspace{-2pt}
\textbf{NVD label:} CWE-787 (Out-of-bounds \textbf{Write})

\textbf{NVD description.} ``heap buffer overflow in
\texttt{get\_one\_sourceline} \dots\ prior to 8.2.4647'' --- generic
overflow wording, no operation direction.

\textbf{AddressSanitizer output (PoC).}
\vspace{-2pt}
\begin{Verbatim}[fontsize=\scriptsize]
ERROR: AddressSanitizer: heap-buffer-overflow ...
READ of size 1 at 0x612000004c6c thread T0
  #0 strlen
  #1 get_one_sourceline scriptfile.c:1930
\end{Verbatim}
\vspace{-2pt}
\textbf{Fix commit (vim 8.2.4647, \texttt{scriptfile.c}).}
\vspace{-2pt}
\begin{Verbatim}[fontsize=\scriptsize]
patch 8.2.4647:"source" can read past end of copied line
  Solution: Add a terminating NUL.
+    if (ga_grow(&ga, 1) == FAIL)
+        break;
+    buf[ga.ga_len++] = NUL;
\end{Verbatim}
\vspace{-2pt}
\hrule
\textbf{Inconsistency.} The sanitizer reports a \hl{READ} past the buffer
in \texttt{strlen}, and the fix adds a terminating NUL to stop the
over-read; NVD assigned the out-of-bounds \emph{write} subtype
(CWE-787). The evidence-supported label is \textbf{CWE-125 (Out-of-bounds
Read)} --- correct family, wrong memory-operation subtype.
\vspace{-4pt}
\end{tcolorbox}
\vspace{-0pt}
\caption{Sub-type mislabel: sanitizer and fix show an out-of-bounds read
(CWE-125), but NVD assigned CWE-787.}
\vspace{-0pt}
\label{fig:case-subtype}
\end{figure}

\textbf{Sub-type mislabel within a family (21.3\%).}
The second pattern keeps the correct family but selects the wrong
sibling, concentrated in memory-safety where an out-of-bounds read is
labeled as a write (CVE-2022-1160, Figure~\ref{fig:case-subtype}). A
structural driver is available: MITRE's View-1003, which NVD uses for
normalization, lacks CWE-122 and maps it to its parent CWE-787, so a
heap over-\emph{read} can be drawn toward the write sub-type because the
read/write distinction is what normalization to a coarser node. This operates within a family, matching the family-level
finding (\S\ref{sec:cwe-dist}) of imprecise sub-types but rare
inconsistency.

\begin{figure}[t]
\begin{tcolorbox}[left=3pt,right=3pt,top=2pt,bottom=3pt,
  lifted shadow={1mm}{-2mm}{3mm}{0.1mm}{black!50!white},
  arc=0pt,auto outer arc,boxrule=.5pt,leftrule=2pt,
  title={\textbf{Case 3 (Injection sink confusion): CVE-2026-28211}},
  fonttitle=\footnotesize, fontupper=\scriptsize]
\vspace{-2pt}
\textbf{NVD label:} CWE-943 (Data Query Logic injection)

\textbf{NVD description.} ``\dots Python expressions embedded in the log
may be \hl{evaluated} \dots\ a maliciously crafted log file can lead to
arbitrary code execution.'' --- the sink is a Python evaluator, with no
data query involved.

\textbf{Fix commit (\texttt{logReader.py}).}
\vspace{-4pt}
\begin{Verbatim}[fontsize=\scriptsize]
- seq = eval(txtSeq)
+ seq = generateSpeechSequence(txtSeq)
\end{Verbatim}

\hrule
\textbf{Inconsistency.} The fix replaces a Python \texttt{eval()} on
attacker-controlled log content with a non-evaluating parser; the
weakness is \hl{eval injection}, i.e.\ \textbf{CWE-95 (Eval Injection)},
a child of code injection (CWE-94). NVD's \textbf{CWE-943} instead names
the \emph{data-query} injection family (SQL/NoSQL) --- the wrong
interpreter, since no query language is involved.
\vspace{-4pt}
\end{tcolorbox}
\vspace{-2pt}
\caption{Injection-sink confusion: the fix removes a Python
\texttt{eval()} (CWE-95), but NVD assigned CWE-943.}
\vspace{-0pt}
\label{fig:case-injection}
\end{figure}

\textbf{Injection sink / interpreter confusion (19.5\%).}
The third pattern keeps the idea that untrusted input reaches a dangerous
sink but names the wrong interpreter: in CVE-2026-28211
(Figure~\ref{fig:case-injection}) a Python \texttt{eval()} (CWE-95) is
labeled as data-query injection (CWE-943). The record is also marked
un-scheduled for enrichment, so the CWE was never
revisited---one way an imprecise sink choice persists.

\begin{figure}[tp]
\begin{tcolorbox}[left=3pt,right=3pt,top=2pt,bottom=3pt,
  lifted shadow={1mm}{-2mm}{3mm}{0.1mm}{black!50!white},
  arc=0pt,auto outer arc,boxrule=.5pt,leftrule=2pt,
  title={\textbf{Case 4 (Access-control / identity conflation): CVE-2024-25618}},
  fonttitle=\footnotesize, fontupper=\scriptsize]
\vspace{-2pt}
\textbf{NVD label:} CWE-306 (Missing Authentication for Critical
Function)

\textbf{NVD description.} ``\dots\ allows new identities from configured
authentication providers (CAS, SAML, OIDC) to \hl{attach to existing
local users with the same e-mail address} \dots\ this results in a
possible account takeover \dots\ [the system] checks the e-mail address
passed by the provider to find an existing account.''

\hrule
\textbf{Inconsistency.} The external provider \emph{does} authenticate
the user; the flaw is that the authenticated identity is bound to a
pre-existing local account by e-mail address alone, enabling takeover.
NVD (NIST) assigned \textbf{CWE-306 (Missing Authentication)}, but
authentication is not missing. The evidence supports \hl{improper
authentication}, \textbf{CWE-287}, which is in fact the assigning CNA's
label; the two sources disagree and the evidence favors CWE-287.
\vspace{-4pt}
\end{tcolorbox}
\vspace{-2pt}
\caption{Access-control/identity conflation: the provider authenticates
the user, so the flaw is improper binding (CWE-287), not missing
authentication (CWE-306).}
\vspace{-10pt}
\label{fig:case-accesscontrol}
\end{figure}

\textbf{Access-control \& identity conflation (15.3\%).}
The fourth pattern confuses the kind of access-control failure---most
often authentication with authorization. In CVE-2024-25618
(Figure~\ref{fig:case-accesscontrol}), the provider does authenticate the
user, so the flaw is improper identity binding (CWE-287), not missing authentication (CWE-306). The two
sources disagree and the evidence favors the CNA.

\textbf{Discouraged or wrong-branch label (10.3\%).}
The fifth pattern uses a CWE MITRE discourages---a high-level category or
``business logic'' bucket---in place of the specific weakness. Merely
abstract uses are usually defensible and were excluded in review; the
confirmed instances are also on the \emph{wrong branch}. In CVE-2023-0565
(Figure~\ref{fig:case-discouraged}) the prohibited category CWE-840 is
assigned to what the patch shows is input validation (CWE-1287),
not even an ancestor of the true weakness---the non-defensible tail of
the discouraged-label behavior of \S\ref{sec:cwe-dist}.

\begin{figure}[h]
\begin{tcolorbox}[left=3pt,right=3pt,top=2pt,bottom=3pt,
  lifted shadow={1mm}{-2mm}{3mm}{0.1mm}{black!50!white},
  arc=0pt,auto outer arc,boxrule=.5pt,leftrule=2pt,
  title={\textbf{Case 5 (Discouraged / wrong-branch label): CVE-2023-0565}},
  fonttitle=\footnotesize, fontupper=\scriptsize]
\vspace{-2pt}
\textbf{NVD label:} CWE-840 (Business Logic Errors).

\textbf{NVD description.} ``Business Logic Errors in GitHub repository
froxlor/froxlor prior to 2.0.10.'' --- merely restates the category
name and gives no mechanism.

\textbf{Fix commit (\texttt{admin\_templates.php}).}
\vspace{-4pt}
\begin{Verbatim}[fontsize=\scriptsize]
+ if (!array_key_exists($language, $languages)) {
+   Response::standardError('templatelanguageinvalid');
+ }
\end{Verbatim}

\hrule
\textbf{Inconsistency.} The fix rejects a user-supplied \texttt{language}
value not in the known set --- a \hl{missing input validation}, not a
business-rule flaw. NVD's \textbf{CWE-840} is a prohibited category and
on the wrong branch; the enabling weakness is CWE-1287.
\vspace{-4pt}
\end{tcolorbox}
\caption{Discouraged, wrong-branch label: the patch shows Improper
Validation of Specified Type of Input (CWE-1287), but NVD assigned the
prohibited category CWE-840.}
\label{fig:case-discouraged}
\end{figure}

\textbf{Multi-CWE accumulation (7.8\%).}
The final pattern is record-level: several CWEs coexist, at least one
correct and at least one unsupported, with the unsupported never removed.
In CVE-2024-25639 (Figure~\ref{fig:case-multicwe}) an NVD-added CWE-77
(command injection) persists beside the CNA's correct XSS labels
(CWE-79/80). The mechanism is additive---labels from different sources
coexist with no reconciliation step to drop those the evidence does not
support.

\begin{figure}[h]
\begin{tcolorbox}[left=3pt,right=3pt,top=2pt,bottom=3pt,
  lifted shadow={1mm}{-2mm}{3mm}{0.1mm}{black!50!white},
  arc=0pt,auto outer arc,boxrule=.5pt,leftrule=2pt,
  title={\textbf{Case 6 (Multi-CWE accumulation): CVE-2024-25639}},
  fonttitle=\footnotesize, fontupper=\scriptsize]
\vspace{-2pt}
\textbf{NVD label:} CWE-77 (Command Injection), CWE-79 (Cross-site
Scripting), CWE-80 (Basic XSS)

\textbf{NVD description.} ``\dots\ clients inadequately sanitize the AI
model's response and user inputs. This can trigger \hl{Cross Site
Scripting (XSS)} via Prompt Injection from untrusted documents \dots\
read by [the application] from the internet.''

\hrule
\textbf{Inconsistency.} Untrusted content is rendered without
sanitization, injecting script into the page --- \hl{cross-site
scripting}; no OS or shell command is involved. \textbf{CWE-79} (XSS) and
its child \textbf{CWE-80} are both correct, but the third label,
\hl{CWE-77 (Command Injection)}, has no support in the evidence: an
\hl{unsupported extra} CWE NVD added during enrichment.
\vspace{-4pt}
\end{tcolorbox}
\caption{Multi-CWE accumulation: the evidence supports only XSS
(CWE-79/80); CWE-77 is an unsupported label.}
\label{fig:case-multicwe}
\end{figure}

The residual 3.9\% of confirmed mislabels involve file- and
path-handling weaknesses that do not fall cleanly into the six patterns
and are too few to characterize as a pattern of their own. Across all six, the assigned CWE is reachable from the record's
\emph{description} or \emph{end-state impact}, while the
evidence-supported weakness is reachable only from the \emph{patch} or
mechanism---so mismatches concentrate where the taxonomy offers an
easier, coarser, or differently-sourced label than the patch warrants.

\vspace{-4pt}
\begin{tcolorbox}[left=1pt,right=1pt,top=0pt, bottom=2pt, lifted shadow={1mm}{-2mm}{3mm}{0.1mm}{black!50!white},arc=0pt,auto outer arc, boxrule=.5pt,leftrule=2pt]
\textbf{Finding 4:} Confirmed mislabels follow six recurring patterns,
led by confusing a weakness with its downstream consequence and
confusing sibling sub-types within a family. All six arise from how the
CWE taxonomy relates labels---cause vs.\ effect, sibling types---rather
than from project- or language-specific mistakes.
\end{tcolorbox}

\begin{figure}[tp]
\centering
\vspace{0pt}
	\includegraphics[width=0.98\linewidth]{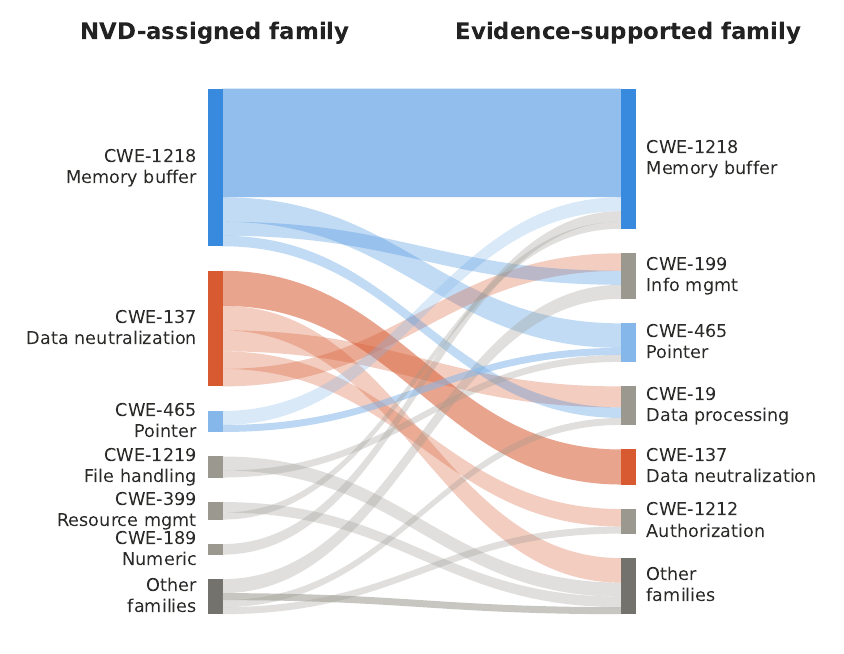}
        \vspace{0pt}
        \caption{Family-level migration of confirmed mislabels.}
\label{fig:effect-chain}
        \vspace{-0pt}
\end{figure}

\vspace{-0pt}
\subsubsection{Failure Effect Chain}
\label{sec:wild:effect-chain}

We now ask
whether the confirmed mislabels, viewed as migrations from the
NVD-assigned CWE to the evidence-supported CWE, concentrate into a few
 confusions or spread broadly across the taxonomy. Each of the
434 confirmed mislabels is a directed pair (NVD label
$\rightarrow$ correct CWE); both endpoints are human-validated, so the
migration is exact rather than model-estimated.

\textbf{Mislabeling is diffuse, not a few recurring swaps.}
The 434 confirmed mislabels comprise 384 distinct (NVD~$\rightarrow$
correct) pairs, nearly all occurring once; the ten most frequent cover
only 11.5\% of the set. One pair stands out: out-of-bounds
write~$\rightarrow$ read (CWE-787~$\rightarrow$ CWE-125), the sub-type
confusion of Case~2 (Figure~\ref{fig:case-subtype}) and the only
recurring swap. Otherwise mislabels do not funnel through a few CWE-pair
confusions but spread broadly across the taxonomy, not a handful of
fixable pairs.

\textbf{Most corrections cross families.}
Mapping each pair to their CWE-699 category shows that
mislabels rarely stay within a weakness family: 80.5\% of
confirmed mislabels migrate \emph{across} families, with the
evidence-supported CWE belonging to a different category than
the NVD label rather than a sibling within the same one. The minority
that stay in-family are dominated by the memory-safety families---the
Memory-Buffer-to-Memory-Buffer band in Figure~\ref{fig:effect-chain}
(driven by the CWE-787$\rightarrow$CWE-125 swaps) is the  heaviest
flow. The remaining flows fan widely: data-neutralization labels
migrate to data-processing, authorization, information-management, and
business-logic families; memory and pointer labels migrate outward to
information-management and resource families. %The breadth confirms the pair-level picture at the family level.

\vspace{-4pt}
\begin{tcolorbox}[left=1pt,right=1pt,top=0pt, bottom=2pt, lifted shadow={1mm}{-2mm}{3mm}{0.1mm}{black!50!white},arc=0pt,auto outer arc, boxrule=.5pt,leftrule=2pt]
\textbf{Finding 5:} Mislabels are diffuse, not a few recurring swaps:
nearly all CWE-pair confusions occur once, and 80.5\% cross weakness
families---so a mislabel usually names a fundamentally different kind of
weakness than the evidence supports, not a sibling of it.
\vspace{-4pt}
\end{tcolorbox}

% =====================================================================
\vspace{-0pt}
\subsection{Other Bad Labeling Practices}
\label{sec:wild:bad-practices}
\vspace{-0pt}
% =====================================================================

\subsubsection{Ambiguous Labels}
\label{sec:wild:ambiguous}
Beyond outright inconsistency, the largest gap between NVD's labels and
a strict standard is \emph{ambiguity}. We quantified its prevalence in
\S\ref{sec:wild:prevalence} (the S2 and S3 outcomes, roughly 31\% of the
corpus); here we treat it as a labeling \emph{practice} problem. Such a
label is defensible but not the most-specific assignment the evidence
supports---not wrong, yet not precise without the consumer resolving the
actual weakness. These are not labeler errors but a consequence of
genuine taxonomy overlap that NVD's records do not flag. They outnumber the evidence-inconsistent mislabels of \S\ref{sec:wild:failure-patterns} by roughly an order of magnitude, yet still impose a cost: each shifts the work of identifying the true weakness onto the consumer.

\vspace{-0pt}
\subsubsection{Discouraged and Prohibited Labels}
\label{sec:wild:disallowed}
A second, distinct practice problem is the use of CWEs that MITRE's own
mapping guidance places off-limits for direct assignment.
MITRE explicitly discourages or prohibits direct use of a subset of
CWEs for CVE labeling, instructing labelers to assign a more-specific
child CWE where possible~\cite{mitre-rcm-guidance}. Despite this
guidance, such labels account for over 16\% of all CWE assignments in
the corpus (2{,}536 CVEs). The five most common---CWE-20, CWE-200,
CWE-400, CWE-284, and CWE-119---are abstract pillar- or class-level
CWEs that can never be the most-specific label for an evidence-supported
weakness. Their prevalence is a systematic departure from MITRE's guidance, not an artifact of ambiguity: each has an S4 rate at or below the corpus average, so these labels are seldom wrong but maximally imprecise.

\vspace{-0pt}
\begin{tcolorbox}[left=1pt,right=1pt,top=0pt, bottom=2pt, lifted shadow={1mm}{-2mm}{3mm}{0.1mm}{black!50!white},arc=0pt,auto outer arc, boxrule=.5pt,leftrule=2pt]
\textbf{Finding 6:} Two labeling practices impose cost without being
outright wrong: ambiguous labels (roughly a third of the corpus) that
are defensible but not most-specific, and discouraged or prohibited
CWEs (over 16\% of assignments) that can never be most-specific. Both
push the work of resolving the true weakness onto the consumer.
\vspace{-4pt}
\end{tcolorbox}

% =====================================================================
\vspace{-6pt}
\subsection{Ecosystem Side Effects}
\label{sec:wild:ecosystem}
\vspace{-6pt}
Beyond the population-level patterns above, {\tech} surfaces three
rarer failure modes that we report qualitatively---as evidence that
mislabeling interacts with the broader ecosystem, not as measured
rates.

\vspace{-6pt}
\subsubsection{Across Label Sources}
\label{sec:wild:cross-source}
A CVE record frequently carries CWE labels from more than one source---an
initial CNA label and an NVD analyst's enrichment---and when these
disagree, the record exposes a mislabel whose \emph{correct} value is
already present elsewhere on it: the error is a failure to reconcile
sources, not a gap in information. This is the instance-level counterpart
to \S\ref{sec:cwe-dist}, where the assigning CNA was the strongest
predictor of label quality.
Two confirmed cases show this directly: CVE-2024-25618
(Figure~\ref{fig:case-accesscontrol}), where the CNA's evidence-supported
CWE-287 and NVD's contradicted CWE-306 coexist, and CVE-2024-25639
(Figure~\ref{fig:case-multicwe}),
where an NVD-added CWE-77 persists beside the CNA's correct XSS labels. In both, the correct label is already on the record; the mislabel is an
unreconciled disagreement.

% \vspace{-4pt}
% \begin{tcolorbox}[left=1pt,right=1pt,top=0pt, bottom=2pt, lifted shadow={1mm}{-2mm}{3mm}{0.1mm}{black!50!white},arc=0pt,auto outer arc, boxrule=.5pt,leftrule=2pt]
% \textbf{Finding 14:} A CVE's CWE labels can come from multiple sources
% (the CNA and NVD enrichment), and when they disagree the
% evidence-supported label is sometimes already present on the record but
% not the one NVD surfaces. Mislabeling here is a failure to reconcile
% co-existing labels, not a lack of information.
% \end{tcolorbox}

\vspace{-6pt}
\subsubsection{Across Labeling History}
\label{sec:wild:cross-history}
A CVE's CWE labels are not fixed at disclosure: NVD records a change
history, and labels are added or revised by later enrichment. That
history is an audit trail, and it sometimes shows an incorrect label
entering a record \emph{after} a correct one was already present---the
error is introduced by a revision later.
CVE-2024-42477 (Figure~\ref{fig:case-history}) is an example:
the evidence-supported CWE-125 (out-of-bounds read) was present at
publication, and enrichment later \emph{added} an unsupported CWE-401
(memory leak) that the record has carried since. The error is thus
traceable to a specific revision rather than the original report. NVD
preserves this provenance, but downstream consumers, who see the
current state, do not.

\begin{figure}[h]
\begin{tcolorbox}[left=3pt,right=3pt,top=2pt,bottom=3pt,
  lifted shadow={1mm}{-2mm}{3mm}{0.1mm}{black!50!white},
  arc=0pt,auto outer arc,boxrule=.5pt,leftrule=2pt,
  title={\textbf{Case 7 (Label introduced by later revision): CVE-2024-42477}},
  fonttitle=\footnotesize, fontupper=\scriptsize]
\vspace{-2pt}
\textbf{NVD label:} CWE-401 (Memory Leak), CWE-125 (Out-of-bounds Read)

\textbf{NVD description.} ``\dots\ the unsafe \texttt{type} member in the
\texttt{rpc\_tensor} structure can cause \hl{global-buffer-overflow}...''

\textbf{Change history.}
\vspace{-4pt}
\begin{Verbatim}[fontsize=\scriptsize]
2024-08-12 record published + CWE-125 (OOB read)
2024-08-15 enrichment analysis + CWE-401 (memory leak)
2026-04-27 later revision (no CWE change)
\end{Verbatim}

\hrule
\textbf{Inconsistency.} An unchecked \texttt{type} field lets a read run
past the tensor buffer, leaking memory --- an \hl{out-of-bounds read}, no
unreleased memory. The evidence-supported \textbf{CWE-125} was present at
publication; enrichment later \hl{added} \textbf{CWE-401 (Missing Release
of Memory)}, a resource-leak weakness the evidence does not support. The
error is traceable to a specific later revision.
\vspace{-4pt}
\end{tcolorbox}
\caption{Label introduced by a later revision: CWE-125 was present at
publication; an unsupported CWE-401 was added during later enrichment.}
\label{fig:case-history}
\end{figure}

% \vspace{-4pt}
% \begin{tcolorbox}[left=1pt,right=1pt,top=0pt, bottom=2pt, lifted shadow={1mm}{-2mm}{3mm}{0.1mm}{black!50!white},arc=0pt,auto outer arc, boxrule=.5pt,leftrule=2pt]
% \textbf{Finding 15:} A record's CWE labels evolve over time, and the
% change history can show an incorrect label being introduced by a later
% enrichment revision rather than by the original report. In
% CVE-2024-42477 the evidence-supported CWE-125 was present at disclosure
% and an unsupported CWE-401 was added three days later, persisting on the
% record thereafter.
% \end{tcolorbox}
\vspace{-6pt}
\subsubsection{Misinformation in Source Reports}
\label{sec:wild:misinformation}
A distinct failure mode originates one step upstream of labeling: the
source report itself---the advisory or description the labeler works
from---omits the mechanism, so a label faithfully derived from it is
still wrong. CVE-2022-1160 (Figure~\ref{fig:case-subtype}) is a verified
instance: its description reports a directionless ``heap buffer
overflow,'' while only the AddressSanitizer output and fixing commit
reveal an out-of-bounds \emph{read}, so the assigned write label
(CWE-787) follows from the report's omission, not a labeler's misreading.
The remedy is upstream---descriptions that record the operation, not just
the weakness class.

\vspace{-4pt}
\begin{tcolorbox}[left=1pt,right=1pt,top=0pt, bottom=2pt, lifted shadow={1mm}{-2mm}{3mm}{0.1mm}{black!50!white},arc=0pt,auto outer arc, boxrule=.5pt,leftrule=2pt]
\textbf{Finding 7:} Mislabeling interacts with the broader ecosystem in
three ways: the correct label is sometimes already on the record from a
disagreeing source, an incorrect label is sometimes introduced by a
later enrichment revision, and some mislabels originate upstream in a
source report that omits the mechanism. In each, the error is one of
reconciliation or provenance, not missing information.
\end{tcolorbox}

% \vspace{-4pt}
% \begin{tcolorbox}[left=1pt,right=1pt,top=0pt, bottom=2pt, lifted shadow={1mm}{-2mm}{3mm}{0.1mm}{black!50!white},arc=0pt,auto outer arc, boxrule=.5pt,leftrule=2pt]
% \textbf{Finding 16:} Some mislabels originate in the source report
% rather than the labeling step: when an advisory or description omits the
% mechanism, a label faithfully derived from it is still wrong. In
% CVE-2022-1160 the description reports a directionless ``heap buffer
% overflow'' while the patch shows an out-of-bounds read, so the assigned
% write label (CWE-787) follows from the report's omission, not a
% labeler's misreading.
% \end{tcolorbox}

\vspace{-0pt}
\section{Discussion}
\label{sec:discussion}
\vspace{-0pt}

%\subsection{What the Labels Mean}
\subsection{Takeaways and Implications}
\label{sec:disc-meaning}
\vspace{-0pt}
The central takeaway for anyone using NVD CWE data is that the dataset
has a large \emph{precision} problem but only a small \emph{correctness}
one, and conflating the two misreads it by an order of magnitude. Most
non-exact labels (the S2/S3 cases, roughly a third of the corpus) are
not errors but defensible consequences of genuine taxonomy overlap---the
same vulnerability admits more than one valid CWE, and NVD frequently
records a broader or sibling one. Only the S4 tail (3.63\%) carries labels that no
reading of the evidence supports. The accurate reading keeps precision and correctness separate: NVD CWE
labels are usually \emph{defensible}, often \emph{imprecise}, and
occasionally \emph{wrong}. 

\vspace{-0pt}
\subsection{Implications for Using NVD CWE Data}
\label{sec:disc-implications}
\vspace{-0pt}
Our claims describe the records and the taxonomy, not the people or process that produce labels; we make no claim about why any individual label was assigned. The label issues nonetheless have direct consequences for the many downstream uses of NVD CWEs---training and evaluating classifiers, building datasets, prioritizing remediation, and measuring trends.

\textbf{Consume the noise as structured, not random.} The labels are most trustworthy at the weakness-\emph{family} level and least at the specific sub-type within it: \S\ref{sec:cwe-dist} shows evidence-inconsistency is low and roughly flat across families while precision varies widely within them. A consumer who needs only ``memory-safety issue'' can rely on the label, but one needing the exact read/write operation often cannot. Because neither the imprecision nor the S4 contradictions are visible from the label alone---both require the patch or advisory to detect---research scored by exact agreement with NVD is measuring agreement with an imprecise source and should account for that imprecision rather than charging all disagreement to the method.

\textbf{Provenance is a free confidence signal.} Label quality varies far more by the assigning CNA than by any intrinsic property of the vulnerability such as severity, and the record-level cases make the mechanism visible---disagreeing NIST- and CNA-supplied labels where the evidence favors one, or an unsupported label added during later enrichment. Since source metadata is on every NVD entry at no cost, weighting or flagging labels by provenance is an immediate improvement over treating all as equally reliable.

\textbf{Patch-grounded auditing is feasible at scale.} That these mismatches are detectable at all is itself a finding: grounding each candidate CWE in the fixing patch rather than the description surfaces exactly the sub-type errors, consequence-for-cause cases, and unsupported co-assigned labels that a description-driven label misses, over the full corpus. We do not argue automated auditing should replace human labeling, only that patch-grounded analysis recovers the information the mismatches turn on, and scales.

\vspace{-0pt}
\subsection{Application to Mitigating Labeling Backlog}
\label{sec:disc-backlog}
\begin{figure}[t]
\centering
\includegraphics[width=0.65\linewidth]{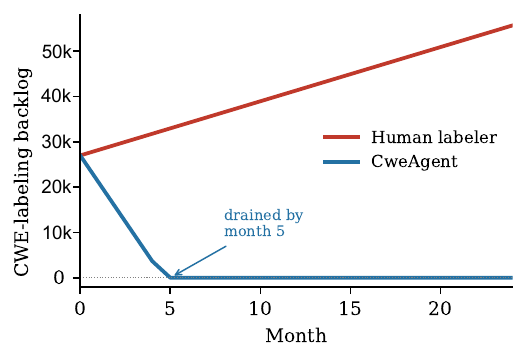}
\vspace{-0pt}
\caption{CWE-labeling backlog under a human labeler versus {\tech}. Arrivals are
approximated at the 2025 average rate (3{,}500/month)~\cite{nistGrowth2026} and each
worker is given a 160~h monthly budget. The model isolates the CWE-labeling subtask and does not
represent full NVD enrichment.}
\vspace{-0pt}
\label{fig:backlog}
\end{figure}

\vspace{-0pt}
As a secondary application, we estimate whether {\tech} could help
reduce the CWE-labeling component of the enrichment backlog. We do not
model the full NVD process; we isolate the subtask {\tech} directly
supports. In a small controlled study, three security-knowledgeable
labelers each timed labeling CVEs sampled from our dataset across four
difficulty bands, given the same evidence {\tech} uses (description,
advisory, patch, code); the median human time was 4.2~min/CVE (mean
4.0). For each month $t$ we treat NVD CVE publications as arrivals
$A_t$ and update the backlog as
$B_t=\max(0,\,B_{t-1}+A_t-C_t)$, comparing one human labeler against one
{\tech} instance under the same monthly time budget, with $C_t$ set by
the measured per-CVE times ({\tech}: 61.7~s; human: 251~s). Starting
from the end-2025 backlog of $\sim$27{,}000 un-enriched
CVEs~\cite{nvdBacklogIG2026} and arrivals at the 2025 average
rate~\cite{nistGrowth2026}, Figure~\ref{fig:backlog} shows that a single
human worker cannot keep pace---its backlog grows---while one {\tech}
instance drains the labeling backlog within a few months. This
illustrates that {\tech} can substantially increase CWE-labeling
throughput and may serve as a candidate-enrichment aid under backlog
pressure, while still requiring human oversight for final NVD-quality
enrichment.

% \subsection{Threats to Validity}
% \label{sec:disc-threats}

% \textbf{Internal validity.} Our headline error figure depends on the
% {\tech} Arbitrator, which operationalizes ``inconsistent with the
% evidence'' as a judgment by fixed models at temperature zero; a faulty
% instrument would distort the S3 rate. We bound this two ways: the
% Arbitrator's reliability is measured directly against a labeled test set
% (\S\ref{sec:rq2}), and the entire S3 population was re-examined by hand,
% so the qualitative findings rest on the 434 manually confirmed mislabels
% rather than the automated flag alone. A second threat is uneven evidence:
% our hierarchy treats the fixing patch as most authoritative, so records
% with an unavailable or partial patch carry weaker evidence and a few
% borderline cases may be misclassified---precisely the cases the manual
% review targets. To avoid inflating any rate, records {\tech} cannot judge
% are excluded from all denominators.

% \textbf{External validity.} Our corpus is open-source CVEs with locatable
% fixing commits and is skewed toward C/C++ memory-safety vulnerabilities,
% so the precision and correctness \emph{rates} we report may not transfer
% to closed-source advisories, underrepresented weakness classes, or
% databases other than NVD. To mitigate it, the six failure \emph{patterns}
% are defined by CWE-to-CWE relationships rather than by our corpus, so
% they should generalize better.

\vspace{-0pt}
\subsection{Threats to Validity}
\label{sec:disc-threats}
\vspace{-0pt}
One main threat is instrument validity: the reported S4 rate depends on {\tech}'s judgments. We mitigate this by evaluating the Classifier and Arbitrator separately %(\S\ref{sec:rq1}--\S\ref{sec:rq2}) 
and by manually reviewing all S4 cases; the strongest claims therefore rest on the 434 manually confirmed mislabels, not automated flags alone. Evidence incompleteness is another threat: advisories and patches vary in detail, and some root causes are only partially exposed. We exclude cases {\tech} cannot judge from the relevant denominators.

Our corpus covers open-source CVEs with locatable fixing commits, so the precision and correctness \emph{rates} may not generalize to closed-source or advisory-only vulnerabilities. Because {\tech} uses LLMs, data leakage is also possible; we reduce this risk by using pre-CWE-assignment CVE records, Wayback snapshots when available, and a post-cutoff benchmark subset in RQ1.

\vspace{-0pt}
\section{Related Work}
\label{sec:related}
\vspace{-0pt}
\textbf{CWE label prediction.}
Prior work predicts such labels from CVE descriptions using hierarchical transformers, encoder models, and LLM-based classifiers~\cite{cve2cwe2024,threatzoom,cve2cwe,ctibench,llama3.1,roberta-cwe}. A related line predicts weakness types from code or fixing diffs, including TreeVul~\cite{pan2023fine}, VulExplainer~\cite{vulexplainer}, and hierarchy-aware classifiers~\cite{ji-emnlp}, often addressing severe CWE class imbalance with re-weighting or margin-based losses~\cite{livable,ldam,cb-loss,logit-adjustment}. These systems generate likely labels. Our goal is different: to audit labels already assigned by NVD. This requires deciding whether a disagreement reflects taxonomy overlap, defensible imprecision, or evidence-inconsistent mislabeling, rather than merely predicting the most likely CWE.

% \textbf{NVD and CWE label quality.}
% Closest to our goal, FixV2W~\cite{fixv2w} flags administratively invalid CVE--CWE mappings, such as assignments to \textsc{Prohibited}, \textsc{Discouraged}, or placeholder CWEs, while broader NVD-quality studies characterize missing, inconsistent, or structurally noisy vulnerability metadata~\cite{cleaning-nvd,threat-kg}. These works expose important metadata-quality problems, but they mainly detect invalidity or inconsistency visible from the record or taxonomy. We study a semantic problem: an administratively valid CWE can still contradict the advisory, patch, or code evidence, while an imprecise label may remain defensible. Detecting this requires evidence-grounded adjudication, not taxonomy checks alone.

\textbf{NVD and CWE label quality.}
Closest to our goal, Anwar et al.'s \emph{Cleaning the NVD}~\cite{cleaning-nvd} systematically studies NVD data-quality problems across publication dates, affected products, severity scores, and vulnerability type information. FixV2W~\cite{fixv2w} further flags administratively invalid CVE--CWE mappings, such as assignments to \textsc{Prohibited}, \textsc{Discouraged}, or placeholder CWEs. These works expose important metadata-quality problems, but mainly detect missing, inconsistent, or rule-invalid information visible from the record or taxonomy. We study a \textit{semantic} problem: an administratively valid CWE can still contradict the advisory, patch, or code evidence, while an imprecise label may remain defensible. Detecting this requires evidence-grounded adjudication, not taxonomy checks alone.

\textbf{LLM agents for security analysis.}
LLMs have been applied to vulnerability detection, localization, repair, and security reasoning~\cite{llm-vuln-prompt,llm-vuln-nier,vuln-survey}. {\tech} uses LLMs for a different measurement role: it composes semantic extraction, full-catalog CWE matching, and label adjudication into an auditing instrument for NVD CWE labels.

\vspace{-0pt}
\section{Conclusion}\label{sec:conclusion}
\vspace{-0pt}
We presented {\tech}, a code-semantics-grounded instrument that audits
NVD CWE labels by comparing each against the weakness its advisory,
patch, and code evidence supports. Measuring 15{,}556 open-source CVEs,
we find NVD labels are usually defensible but often imprecise and only
occasionally evidence-inconsistent (3.63\%), with quality tracking the
assigning CNA and weakness type rather than severity. Our code, data, and
confirmed-mislabel set are available at
\url{https://figshare.com/s/6ba6621a69a2dac5b516}.

\bibliographystyle{IEEEtran}
\bibliography{paper}

\end{document}